\documentclass[aps,amsmath,amssymb,reprint,superscriptaddress]{revtex4-2}

\usepackage{graphicx, xcolor}
\usepackage{dcolumn}
\usepackage{bm}
\usepackage[unicode=true,pdfusetitle,
 bookmarks=true,bookmarksnumbered=false,bookmarksopen=false,
 breaklinks=false,pdfborder={0 0 1},backref=false,colorlinks=false]{hyperref}
\usepackage{placeins}
\usepackage{float}

\begin{document}

\preprint{APS/123-QED}

\title{Dynamical Backaction Magnomechanics}

\author{C.A. Potts}
\affiliation{Department of Physics, University of Alberta, Edmonton, Alberta T6G 2E9, Canada}
\email{cpotts@ualberta.ca, jdavis@ualberta.ca}

\author{E. Varga}
\affiliation{Department of Physics, University of Alberta, Edmonton, Alberta T6G 2E9, Canada}

\author{V.A.S.V. Bittencourt}%
\affiliation{Max Planck Institute for the Science of Light, Staudtstr. 2, PLZ 91058 Erlangen, Germany}%

\author{S. {Viola Kusminskiy}}
\affiliation{Max Planck Institute for the Science of Light, Staudtstr. 2, PLZ 91058 Erlangen, Germany}%
\affiliation{Institute for Theoretical Physics, University Erlangen-Nuremberg, Staudtstr. 7, PLZ 91058 Erlangen, Germany}

\author{J.P. Davis}
\affiliation{Department of Physics, University of Alberta, Edmonton, Alberta T6G 2E9, Canada}

\begin{abstract}

Dynamical backaction resulting from radiation pressure forces in optomechanical systems has proven to be a versatile tool for manipulating mechanical vibrations. Notably, dynamical backaction has resulted in the cooling of a mechanical resonator to its ground-state, driving phonon lasing, the generation of entangled states, and observation of the optical-spring effect. In certain magnetic materials, mechanical vibrations can interact with magnetic excitations (magnons) via the magnetostrictive interaction, resulting in an analogous magnon-induced dynamical backaction. In this article, we directly observe the impact of magnon-induced dynamical backaction on a spherical magnetic sample's mechanical vibrations. Moreover, dynamical backaction effects play a crucial role in many recent theoretical proposals; thus, our work provides the foundation for future experimental work pursuing many of these theoretical proposals. 

\end{abstract}
\maketitle

\section{Introduction}

Hybrid cavity systems hold great promise for exploring a wide variety of physical phenomena.  One broad example of this is the rapid maturation of cavity optomechanics, i.e.~coupling electromagnetic cavities with mechanical degrees of freedom \cite{aspelmeyer2014cavity}. The coupling of electromagnetic cavities with magnonic systems has also generated significant interest \cite{lachance2019hybrid,awschalom2021quantum,li2020hybrid}, with theoretical proposals for magnetometry \cite{ebrahimi2020ultra} and axion detection \cite{crescini2020axion,ikeda2021axion,flower2019broadening}, and experiments demonstrating strong-coupling \cite{huebl2013high,zhang2014strongly,tabuchi2014hybridizing,goryachev2014high,potts2020strong}, magnon Fock state detection \cite{lachance2020entanglement,lachance2017resolving}, coupling to superconducting qubits \cite{tabuchi2015coherent,tabuchi2016quantum}, bidirectional microwave to optical conversion \cite{hisatomi2016bidirectional,zhu2020waveguide}, Floquet electromagnonics \cite{xu2020floquet}, and non-reciprocity \cite{wang2019nonreciprocity}.

\begin{figure}[b]
\includegraphics[width = 0.45\textwidth]{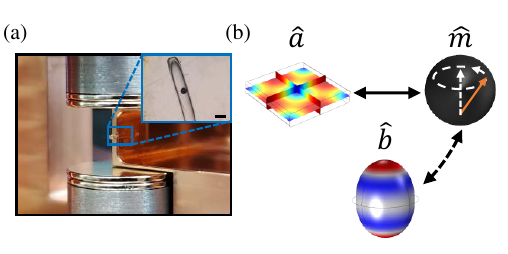}
\caption{Experimental set-up. (a) Photograph of half of the microwave cavity, made from oxygen-free copper. The YIG sphere is placed -- free to move -- within a glass capillary, inner dimension of approximately 300 microns. A set of permanent neodymium magnets attached to a pure iron yoke provide the bias magnetic field, a solenoid (not shown) allows the bias field to be varied dynamically. Inset: Optical micrograph of the YIG sphere inside of the glass capillary. Scale bar is 500 microns.  (b) Schematic of the coupled magnomechanical system. $\hat{a}$ -- Numerical simulation of the $\textrm{TE}_{\textrm{101}}$ microwave magnetic field distribution. $\hat{m}$ -- Schematic representation of the Kittel magnon mode within the spherical YIG sample. $\hat{b}$ --  Numerically simulated displacement of the mechanical mode that has the strongest magnomechanical coupling to the Kittel mode. The solid line indicates linear magnon-photon coupling, while the dashed line represents parametric magnon-phonon coupling.}
\label{Fig:01}
\end{figure}

Combining these concepts yields the field of cavity magnomechanics: coupling of magnetic excitations (magnons) with both an electromagnetic cavity as well as a mechanical resonator \cite{zhang2016cavity}, as illustrated in Fig.~\ref{Fig:01}(b).  Furthermore, an important triple-resonance condition is possible, where the phonon frequency matches the difference in frequencies between the hybrid cavity-magnon modes \cite{zhang2016cavity}, allowing selective cavity enhancement of scattering processes. 
This triple-resonance system has sparked significant interest, resulting in theoretical proposals for the generation of non-classical entangled states \cite{cheng2021tripartite,nair2020deterministic,li2021entangling,li2019entangling,yang2020nonreciprocal,li2018magnon}, squeezed states \cite{li2021squeezing, li2019squeezed,zhang2021generation}, classical and quantum information processing \cite{sarma2021cavity,li2020phase,yang2021nonreciprocal,zhao2021phase,kong2019magnetically}, quantum correlation thermometry \cite{potts2020magnon}, and exploring $\mathcal{PT}$-symmetry  \cite{ding2021enhanced,wang2020magnon,yang2020ground,wang2019magnon}.  Many of these proposals rely on the ability of the external drive to act on the mechanical motion, so called dynamical backaction.  Yet, despite the large number of theoretical papers, there remains to date a single experimental observation of cavity magnomechanics \cite{zhang2016cavity}.

Here, we report cavity magnomechanical detection of all of the fundamental dynamical backaction effects, i.e.~dynamical heating (amplification) and cooling (damping) --- leading to phonon lasing \cite{mahboob2013phonon,ding2019phonon} and noise squashing \cite{rocheleau2010preparation} --- and the first observation of the magnonic spring effect \cite{potts2020magnon}.  These results are enabled, in part, by homodyne detection of the mechanics, with reduced clamping and viscous damping.  Our observations of dynamical backaction in a cavity magnomechanical system open the door for many of the above theoretical proposals, in particular those that involve entanglement \cite{cheng2021tripartite,nair2020deterministic,li2021entangling,li2019entangling,yang2020nonreciprocal,li2018magnon}, squeezed states \cite{li2021squeezing, li2019squeezed,zhang2021generation}, and thermometry \cite{potts2020magnon}.  With reasonable improvements, future experiments could realize, for example, magnon-mediated cooling of the mechanics into the ground-state, which would achieve the largest mechanical system to date to be taken into its quantum ground state, and possibly allow tests of gravitational decoherence currently being pursued with levitated spheres \cite{romero2011quantum,romero2011large,childress2017cavity,delic2020cooling}.

\section{Theoretical Background}

Hybrid magnomechanical systems are composed of three parts: the electromagnetic field (confined in the microwave cavity), the magnetic excitations of a dielectric, and its mechanical vibrations (both hosted in the dielectric YIG sphere). To study the dynamical response of the coupled cavity magnomechanical system, we consider three interacting bosonic modes, as shown in Fig.~\ref{Fig:01}, with annihilation operators $\hat{a}$ (microwave cavity mode), $\hat{m}$ (magnon mode), and $\hat{b}$ (phonon mode), with frequencies $\omega_{\textrm{a}}$, $\omega_{\textrm{m}}$, and $\Omega_{\textrm{b}}$ respectively. It should be noted while we have used a quantum description of the theory, all experimental results presented in the article are classical.

Before tackling the full interacting Hamiltonian, we first study the microwave-magnon interaction. The linear interaction between magnons and microwaves is the hallmark of cavity magnonic systems \cite{huebl2013high,zhang2014strongly,tabuchi2014hybridizing}, and the Hamiltonian describing a magnon mode interacting with a single driven microwave cavity mode reads (in the frame rotating with the driving frequency) \cite{zhang2014strongly,tabuchi2014hybridizing}
\begin{equation}
\begin{aligned}
    \mathcal{H} &= -\hbar\Delta_{\textrm{a}} \hat{a}^{\dagger}\hat{a} - \hbar\Delta_{\textrm{m}} \hat{m}^{\dagger}\hat{m} \\&+ \hbar g_{\textrm{am}}(\hat{a}\hat{m}^{\dagger} + \hat{a}^{\dagger}\hat{m}) + i\hbar\epsilon_{\rm{d}}\sqrt{\kappa_{\rm{ext}} }(\hat{a} - \hat{a}^{\dagger}).
\label{Hamiltonian01}
\end{aligned}
\end{equation}
\noindent The magnon-photon coupling rate is given by $g_{\textrm{am}}$, the external drive is $\epsilon_{\rm{d}} = \sqrt{\mathcal{P}/\hbar \omega_{\rm{d}}}$ (where $\mathcal{P}$ is the external drive power at the device), $\kappa_{\rm{ext}}$ is the external coupling rate, and $\hbar$ is the reduced Planck's constant. The cavity and magnon detunings are defined as $\Delta_{\textrm{a}} = \omega_{\textrm{d}} - \omega_{\textrm{a}}$, and $\Delta_{\textrm{m}} = \omega_{\textrm{d}} - \omega_{\textrm{m}}$ respectively, where $\omega_d$ is the external drive frequency.  The magnon frequency is given by $\omega_m = \gamma \vert \textbf{B}_0 \vert$, where $\gamma/2 \pi = 28$ GHz/T is the gyromagnetic ratio, and $\textbf{B}_0$ is the applied static magnetic field \cite{fletcher1959ferrimagnetic}. The first two terms in the Hamiltonian represent the occupancy of the photon and magnon modes, respectively. The third term describes the linear magnon-photon coupling, and the final term describes the external drive.

\begin{figure}[t]
\includegraphics[width = 0.48\textwidth]{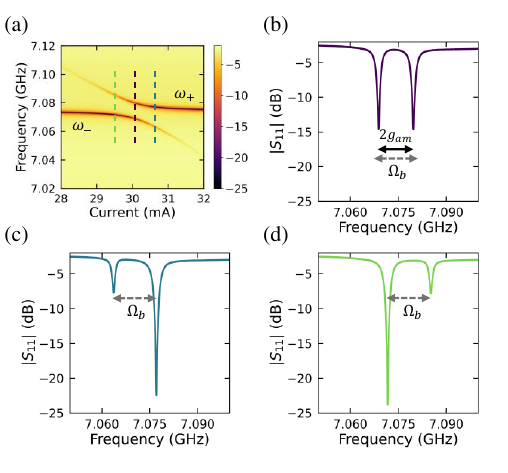}
\caption{Normal mode spectrum. (a) Measured normal mode spectrum as a function of static magnetic field. Dashed lines correspond to the spectrum in (b-d). (b) Cavity reflection spectrum when the magnon is resonant with the bare-cavity mode. The normal mode splitting, $2g_{\textrm{am}}$ is smaller than the phonon frequency, $\Omega_{\textrm{b}} = 2\pi \times 12.6278$ MHz. (c) Cavity reflection spectrum when the magnon frequency is smaller than the bare-cavity resonance frequency. The system was set-up such that the normal mode spacing matches the mechanical frequency. Here the lower mode is `magnon-like' and the upper mode is `photon-like.' (d) Cavity reflection spectrum when the magnon frequency is larger than the bare-cavity resonance frequency, similar to the detuning in (c). Here the upper mode is `magnon-like' and the lower mode is `photon-like.' }
\label{Fig:02}
\end{figure}

Due to the linear coupling, microwaves and magnons hybridize: the normal modes of the interacting Hamiltonian Eq.~\eqref{Hamiltonian01} are superpositions of magnons and photons. We label these modes as $+$ and $-$, and the difference between their frequencies is given by
\begin{equation}
    \omega_+-\omega_- \equiv \Delta\omega = \sqrt{4g_{\textrm{am}}^2 + \Delta_{\textrm{am}}^2},
\end{equation}
\noindent where $\Delta_{\textrm{am}} = \omega_{\textrm{a}} - \omega_{\textrm{m}}$ is the magnon-photon detuning. Since an externally applied bias field can tune the magnon frequency, the hybridization of the modes is controllable: e.g., by varying the current through a solenoid providing the bias field. Furthermore, when the cavity is resonant with the magnon mode, $\omega_{\textrm{a}}=\omega_{\textrm{m}}$, the normal modes are a maximal hybridization of magnons and photons. Otherwise, the normal modes describe partial hybridization; one of the modes is `magnon-like', and the other `photon-like'. The normal mode splitting can be directly measured via the reflected microwave signal. This can be seen in Fig.~\ref{Fig:02}, which depicts the normal mode splitting in our experiment. 

The value of the coupling rate can be extracted by performing a fit to the normal modes presented in Fig.~\ref{Fig:02}(a), using the theory from Ref.~\cite{zhang2014strongly}. For our experimental configuration we extract a coupling rate $g_{\textrm{am}}/2\pi = 5.43$ MHz. Furthermore, both the magnon and microwave modes are subject to decay. The microwave cavity decay rate is composed of both an intrinsic cavity decay and the coupling to external coaxial cables (used to drive/measure the cavity). The main source of magnon damping is intrinsic Gilbert damping \cite{gilbert2004phenomenological}, which includes dissipation processes associated with electron-lattice coupling. Other possible sources of damping include two-magnon scattering processes between the magnetostatic mode and the spin-wave continuum \cite{nemarich1964contribution}. Those processes yield an inhomogeneous broadening of the magnon linewidths for different magnon modes \cite{klinger2017gilbertdamping}, which are less prominent in well-polished spheres. From the measured data we extract the magnon decay rate $\gamma_m/2\pi = 1.01$ MHz, and the total cavity decay rate $\kappa/2\pi = 3.87$ MHz, placing our experiment well within the strong-coupling regime, $g_{\textrm{am}} > \{ \kappa,\gamma_m \}$.

 Besides the coupling to microwave photons, magnons couple to the mechanical vibrations of the material. The magnon-phonon coupling is mediated by the magnetostrictive interaction \cite{keshtgar2014acoustic,zhang2016cavity,callen1968magnetostriction}, which couples magnetization to mechanical strain. The magnetoelastic coupling originates from the effects of lattice strain combined with exchange interactions, dipole-dipole interactions, and spin-orbit coupling \cite{gurevich1996magnetization}. Under strain, the distance between magnetic atoms/ions changes and, as a consequence of the aforementioned interactions, the orbital and spin configurations are modified. The overall effect is a coupling between magnetic order and strain/stress, which is obtained from the magnetic anisotropy energy density \cite{gurevich1996magnetization,keshtgar2014acoustic}. For small vibrations and magnetization oscillations, this energy density term yields two types of magnon-phonon couplings: a linear coupling, relevant for resonant excitations, and a parametric coupling \cite{zhang2016cavity}. Our experiments investigate the later, which has a form analogous to the photon-phonon coupling in optomechanical cavities, where the optical cavity frequency is modulated by mechanical vibrations, for example, of a movable mirror \cite{aspelmeyer2014cavity}. In our experiment, the magnetic element is a YIG sphere, which acts as both the magnon and phonon resonator. YIG is the material of choice for magnomechanical experiments since it possesses excellent magnetic \cite{huebl2013high,zhang2014strongly,tabuchi2014hybridizing} and mechanical properties \cite{spencer1958magneto}. In fact, YIG spheres have mechanical decay rates similar to spheres of silicon and quartz \cite{lecraw1961extremely}. 

The parametric magnetostrictive coupling is described by a radiation-pressure-like parametric Hamiltonian \cite{zhang2016cavity}:
\begin{equation}
    \mathcal{H}_{\textrm{int}} = \hbar g_{\textrm{mb}}^0 \hat{m}^{\dagger}\hat{m}(\hat{b} + \hat{b}^{\dagger}).
    \label{IntHam}
\end{equation}
\noindent Here, the single magnon-phonon coupling strength is given by $g_{\textrm{mb}}^0$. The exact value of the coupling rate depends on the mode overlap between the specific magnon and phonon modes. The lowest order spherical modes are described by $\textrm{S}_{1,l,m}$, where $l$ and $m$ are the angular and azimuthal mode numbers, respectively. For this work, we specifically focus on the spherical $\textrm{S}_{1,2,0}$ mode, shown in Fig.~\ref{Fig:01}(b), since it possesses the largest magnon-phonon coupling rate. We used a commercial YIG sphere, nominally 250 microns in diameter \cite{FerriSphere}. The phonon frequency and magnon-phonon coupling rate can be estimated numerically using COMSOL multiphysics. For the mode of interest, for a 269-micron diameter sphere, the coupling rate is predicted to be $g_{\textrm{mb}}^0 / 2\pi \approx 4.73$ mHz, and the phonon frequency is estimated to be $\Omega_{\rm{b}}/2\pi = 12.66$ MHz. Note, the coupling rate could be increased by decreasing the YIG sphere's diameter; however, this has the effect of decreasing the magnon-photon coupling rate.

In our experiment, the resonant normal mode splitting (for $\omega_{\textrm{a}}=\omega_{\textrm{m}}$) and the phonon frequency are not perfectly matched; indeed, the phonon frequency is slightly larger than the normal mode splitting, Fig.~\ref{Fig:02}(b). However, by detuning the magnon frequency slightly from the bare microwave cavity frequency (i.e. $\Delta_{\textrm{am}} \neq 0$), the normal mode splitting can be tuned to exactly match the phonon frequency, see Fig.~\ref{Fig:02}(c,d). This results in the system becoming triply resonant, which significantly enhances the magnon-phonon coupling.  Indeed, it has been shown in Ref.~\cite{zhang2016cavity} when compared with off-resonance driving the cooperativity of a fully-hybridized triple-resonant system will be enhanced by a factor of
\begin{equation}
    F = 16 \bigg(\frac{\Omega_b}{\kappa+\gamma_m}\bigg)^2.
    \label{F}
\end{equation}

Finally, when the magnon-phonon interaction Hamiltonian Eq.~\eqref{IntHam} is combined with the microwave-magnon Hamiltonian, Eq.~\eqref{Hamiltonian01}, a more complex scenario occurs. Since the magnon-phonon coupling is weak compared to the microwave-magnon coupling, we can describe the magnon mode as a superposition of the normal modes discussed above. The magnon-phonon Hamiltonian can then be understood as describing a scattering process between the normal modes, mediated by a phonon. For example, a drive photon resonant with the higher frequency normal mode is scattered into the lower frequency normal mode and a phonon is generated. Such a process is resonant if the phonon frequency matches the normal mode splitting, fulfilling the triple-resonance condition. The sphere diameter was specifically chosen for our system to operate near the triple-resonance.

\section{Experimental configuration}

The system used in our experiment consists of a three-dimensional microwave cavity machined from oxygen-free high-conductivity copper, seen in Fig.~\ref{Fig:01}(a). The microwave cavity has inner dimensions $30 \times 30 \times 6 \; \textrm{mm}^3$, resulting in the $\textrm{TE}_{\textrm{101}}$ mode having a frequency of $\omega_{\textrm{a}}/2\pi = 7.074$ GHz. The intrinsic decay rate of the microwave cavity is $\kappa_{\textrm{int}}/2\pi = 1.56$ MHz. Coupling to the cavity is achieved through a pair of coaxial cables with external coupling rates of $\kappa_1/2\pi = 1.11$ MHz, and $\kappa_2/2\pi = 1.20$ MHz, resulting in a total cavity linewidth of $\kappa/2\pi = (\kappa_{\textrm{\rm{int}}}+\kappa_{\textrm{1}}+\kappa_{\textrm{2}})/2\pi = 3.87$ MHz.

The single crystal YIG sphere is placed free to move -- to avoid mechanical clamping losses -- within a 300-micron inner diameter capillary \cite{lecraw1961extremely}. The sphere is located near the magnetic field maximum of the microwave cavity and is held in place, and oriented along its easy axis, by the applied static magnetic field. A pair of neodymium magnets provide the static magnetic field, seen in Fig.~\ref{Fig:01}(a). Tunability of the static magnetic field is provided via a $\sim 10^4$ turn solenoid, wrapped around a pure iron core and connected to the permanent magnets using an iron yoke \cite{tabuchi2015coherent}.

To determine the mechanical frequency, and ensure constancy of the magnomechanical coupling rate $g^0_{\textrm{mb}}$ between various measurement techniques, we first performed a series of magnomechanically induced transparency (MMIT) measurements similar to that presented in Ref.~\cite{zhang2016cavity}. From this set of measurements (described in Appendix~\ref{Transparency}) we are able to extract a phonon frequency $\Omega_{\rm{b}} / 2\pi = 12.6270$ MHz, a single magnon-phonon coupling rate of $g^0_{\textrm{mb}} / 2\pi = 4.38$ mHz, and an intrinsic mechanical linewidth of $\Gamma_{\rm{b}}/2\pi = 286$ Hz.

\begin{figure*}[t]
\includegraphics[width = 0.98\textwidth]{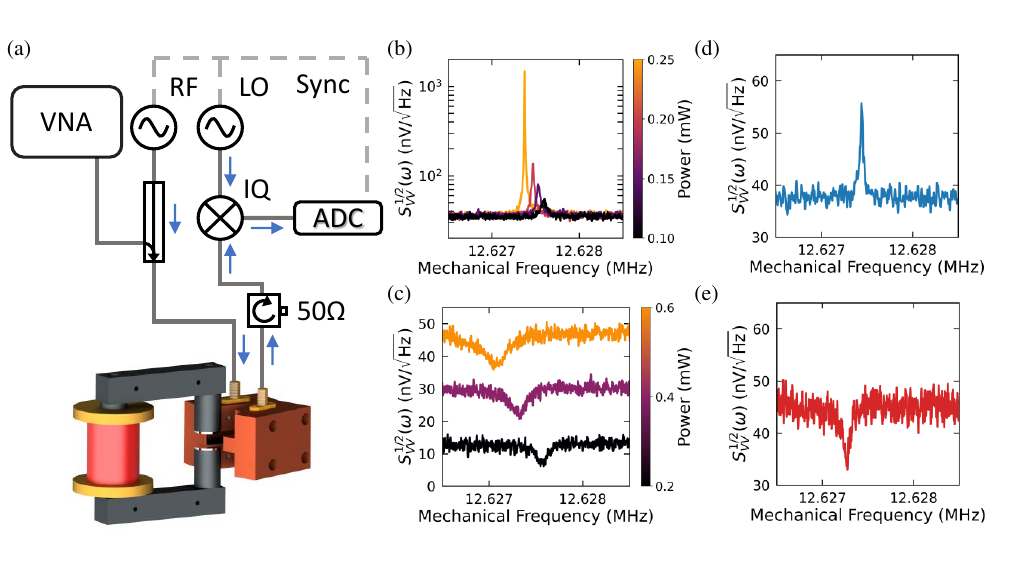}
\caption{Homodyne mechanical detection. (a) Simplified schematic of the measurement setup: VNA, vector network analyzer; RF/LO, microwave generators; ADC, analog-to-digital converter; IQ, IQ-mixer, both the in-phase and quadrature ports were connected to the ADC. (b) Power spectral density of the mechanical motion. The normal mode spectrum was tuned to Fig.~\ref{Fig:02}(d). The probe was tuned directly on resonance with the upper normal mode. With increasing drive power, a frequency shift, and linewidth narrowing can be observed. (c) Power spectral density of the mechanical motion; offset for clarity. The normal mode spectrum was tuned to Fig.~\ref{Fig:02}(c). The probe was tuned directly on resonance with the lower normal mode. Due to interference between scattered excitations and the thermal magnon bath, the power spectrum lies below the noise floor, known as noise squashing. (d) Power spectral density of the mechanical motion, the drive tone was detuned one mechanical frequency above the upper normal mode. (e) Power spectral density of the mechanical motion, the drive tone was detuned one mechanical frequency below the lower normal mode. For both (d) and (e), the drive power was held constant at 12 mW. All data presented was obtained with the experimental setup in a partial pressure of helium gas.}
\label{Fig:04}
\end{figure*}

In the rest of the experiment, we measure the YIG sphere's mechanical vibrations without resorting to the MMIT window, which more easily enables the observation of dynamical backaction effects. In this scheme, a microwave signal was sent to the hybrid magnomechanical system, as shown in Fig.~\ref{Fig:04}(a). The transmitted signal was demodulated using an IQ-mixer, and the low-frequency mechanical signal was digitized using an analog-to-digital (ADC) converter. The reflected signal was passed through a directional coupler and measured using a VNA to characterize the normal mode spectrum. During the mechanical measurements, the VNA was not exciting the cavity to avoid potential beat frequencies from obfuscating the mechanics. To balance the homodyne circuit, the DC component of the demodulated signal was continually measured and locked dynamically -- at a rate of 1 kHz -- by adjusting the local oscillator phase. Data was taken in three atmospheric conditions: ambient pressure, a partial pressure of pure helium gas ($\sim 15$ Torr), and vacuum ($< 1$ mTorr). 

\section{Homodyne Mechanical Detection}

As mentioned, one remarkable aspect of this hybrid magnomechanical system is that it can be brought into triple-resonance. In this case, the red mechanical sideband of the upper normal mode has the same frequency as the lower normal mode and vice versa. Therefore, not only is the drive tone cavity-enhanced, but one of the two mechanical sidebands is simultaneously cavity-enhanced. The Stokes (anti-Stokes) scattering process is strongly preferred, resulting in effective magnomechanical backaction heating (cooling). These two specific scenarios will be discussed in more detail in the next sections. 

The mechanical power spectrum under the triple-resonance condition is shown in  Fig.~\ref{Fig:04}(b,c). The normal mode spectrum for Fig.~\ref{Fig:04}(b) is shown in Fig.~\ref{Fig:02}(d), and the drive is tuned on resonance with the upper normal mode. With increasing drive power, two effects can be seen in Fig.~\ref{Fig:04}(b): the frequency of the mode decreases, we attribute this to parasitic thermal effects that are discussed in Appendix~\ref{Heating}, and a narrowing of the linewidth, which ultimately results in phonon lasing discussed in Section~\ref{Parametric}.  Conversely, for Fig.~\ref{Fig:04}(c), the normal mode spectrum is shown in Fig.~\ref{Fig:02}(c), and the drive is tuned on resonance with the lower normal mode. Again we observe a power dependent frequency shift resulting from parasitic thermal effects. We further see a phenomena known as \textit{noise squashing} resulting from interference between the thermal magnon bath and excitations scattered via the magnomechanical interaction, which will be discussed further in Section~\ref{Squash}.

Furthermore, it is possible to observe mechanical motion without relying on the triple-resonance condition, one can apply the drive tone on the blue sideband of the upper normal mode (or the red sideband of the lower normal mode). Although this sideband driving has some similarities with procedures commonly adopted in cavity optomechanics \cite{aspelmeyer2014cavity}, the composition of the normal modes can be changed by varying the magnon-photon detuning. This in turn results in each normal mode experiencing a detuning-dependent coupling rate, decay rate, and effective phonon coupling rate. Therefore, there exists an optimal detuning for mechanical measurements. The magnon and photon components of the normal mode provide two distinct operations, the magnon-like component couples directly to the mechanical motion, whereas the applied microwave tone can drive the photon component. Thus, the competition between these two effects needs to be balanced for optimal mechanical detection. The mechanical power spectrum while driving above (below) the upper (lower) normal mode, with optimal detunings, are shown in Fig.~\ref{Fig:04}(d) and (e), respectively. Due to the small magnomechanical coupling (compared to the microwave-magnon coupling), a drive power of $\sim 5$ mW was required to resolve the mechanical spectrum. All powers in this article are quoted as power at the device, which were carefully calibrated for each experimental configuration.

It should be noted that unlike many optomechanical measurements, the observed mechanical motion is not thermomechanical in nature. Due to the high drive powers, there is considerable backaction in the form of heating (cooling) of the mechanical mode. However, the intrinsic mechanical properties can be extracted by considering the data presented in Fig.~\ref{Fig:04}(b) and extrapolating to zero drive power. As a result, within a partial helium environment, we find the mechanical mode has a resonance frequency $\Omega_{\textrm{b}} = 12.6278$ MHz, and intrinsic mechanical decay rate $\Gamma_{\textrm{b}}/2\pi = 98$ Hz.

 Finally, to confirm the observed mechanical signal was \textit{not} a result of direct electromechanical coupling (i.e. coupling between photons and phonons), measurements were performed, shifting the magnon frequency far from the microwave resonance frequency. In this scenario, the normal mode spectrum disappears, and we are left with only the microwave cavity resonance. Both OMIT and homodyne measurements were performed; no evidence of the mechanical motion was observed in either of these scenarios. Therefore we are confident the magnon indeed mediates the phonon interaction.

\subsection{Magnon Spring Effect}

\begin{figure}[t]
\includegraphics[width = 0.45\textwidth]{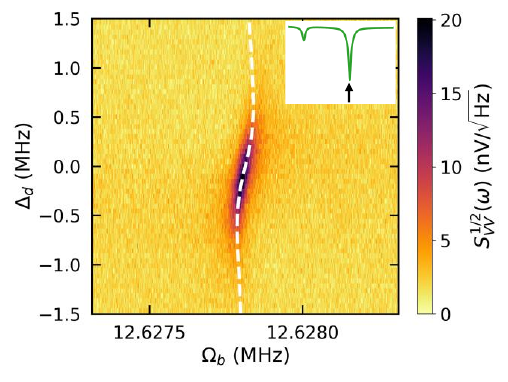}
\caption{Magnon spring effect. Power spectral density of the mechanical motion within a partial pressure of helium gas and a constant probe power of 0.13 mW. The microwave drive was tuned to the normal mode shown in the inset. White curve is the theoretical prediction for the magnon-spring effect \cite{potts2020magnon}, where the only fit parameter was the intrinsic mechanical frequency.}
\label{Fig:06}
\end{figure}

The magnomechanical interaction is given by the interaction Hamiltonian Eq.~\eqref{IntHam}. As seen above, this results in the formation of sidebands that carry information about the mechanical vibrations. However, the interaction also results in a modification of the mechanical susceptibility due to dynamical backaction from the interaction with magnons. We have previously described the full linear theory of the magnomechanical interaction in Ref.~\cite{potts2020magnon} and derived the following expression for the phonon self-energy,
\begin{equation}
    \Sigma[\omega] = i\vert g_{\textrm{mb}}\vert^2 (\Xi[\omega] - \Xi^*[-\omega]).
    \label{SelfEnergy}
\end{equation}
Here, $g_{\textrm{mb}} = g^0_{\textrm{mb}}\langle m \rangle$ is the cavity enhanced magnon-phonon coupling rate, $\vert \langle m \rangle \vert^2$ is the coherent steady-state magnon population, and $\Xi[\omega] = [\chi^{-1}_{\textrm{m}}[\omega] + g^2_{\textrm{am}}\chi_{\textrm{a}}[\omega]]^{-1}$. The magnon and cavity susceptibilities are given by, $\chi_{\textrm{m}}[\omega] = [-i(\Delta_{\textrm{m}}+\omega) + \gamma_{\textrm{m}}/2]$ and $\chi_{\textrm{a}}[\omega] = [-i(\Delta_{\textrm{a}}+\omega) + \kappa/2]$, respectively. In the weak coupling limit, when $g_{\textrm{mb}} \ll {\kappa,\gamma_{\textrm{m}}}$ -- which holds for all data presented in this article and would only break down for the highest on-resonance drive powers -- the real and imaginary parts of the self-energy describe a mechanical frequency shift $\delta\Omega_{\textrm{b}} = -\textrm{Re}\Sigma[\omega]$, the magnon-spring effect, and an additional magnomechanical damping rate $\Gamma_{\textrm{mag}} = 2\textrm{Im}\Sigma[\omega]$.

In order to observe the small magnon-induced frequency shift, parasitic heating needed to be eliminated (see Appendix~\ref{Heating}). To reduce heating of the sphere, a low drive power was required; however, reducing the drive power simultaneously reduces the frequency shift. Therefore, this experimental run was performed in a low pressure ($\sim$ 15 Torr) of pure helium gas to reduce the mechanical linewidth, allowing the small frequency shift to be resolved. Helium was used because it possesses high thermal conductivity; therefore, it provides good thermalization while limiting the mechanical damping of the sphere. Secondly, heating of the sphere was primarily due to magnon decay and not microwave photon absorption. Thus, unlike in Fig.~\ref{Fig:04}(b), the drive tone was applied to the `photon-like' normal mode and the interaction with phonons scattered excitations into the `magnon-like' normal mode. This indeed resulted in less heating of the YIG sphere; however, it has the unwanted secondary effect of reducing the detection efficiency due to the reduced external coupling of the `magnon-like' mode. 

Figure~\ref{Fig:06} shows the mechanical power spectrum, revealing the magnon-induced frequency shift. To avoid complicating effects from heating or a slow-drift of the magnetic field, the magnet was re-adjusted between each drive frequency, and the drive frequencies were applied in a randomized order. This data was taken with a drive power of $0.13$ mW; at this power and drive detuning we observed negligible frequency shift due to heating. The white curve is a theoretical prediction from Eq.~\eqref{SelfEnergy}, where the only fit parameter used was the intrinsic mechanical frequency, which increased slightly in the partial pressure of helium, $\Omega_{\textrm{b}}/2\pi = 12.6278$ MHz. All other parameters were extracted using a fit to the reflected normal mode spectrum and the magnomechanical damping measurement discussed below. The theoretical prediction and the experimentally measured shift are in agreement, confirming the direct observation of the magnon spring effect. 

\subsection{\label{Parametric}Magnomechanical Anti-Damping}

\begin{figure}[t]
\includegraphics[width = 0.48\textwidth]{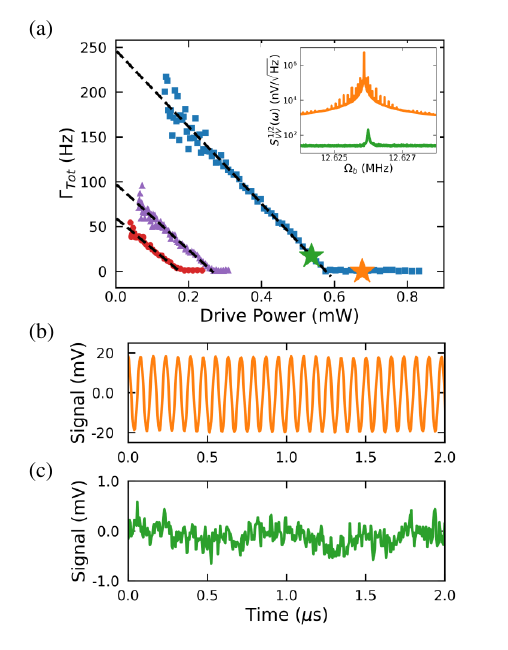}
\caption{Parametric instability. (a) Mechanical linewidth as a function of drive power in air (blue squares), helium partial pressure $\sim 15$ mTorr (purple triangles), and vacuum (red circles). The single magnon-phonon coupling rate extracted from these curves was $g_{\textrm{mb}}^0 / 2\pi = 4.65,\, 4.66,\, 4.43$ mHz, respectively. For this data the normal mode spectrum was tuned to (d) in Fig.~\ref{Fig:02}, and the probe tone was tuned on resonance with the `magnon-like' mode. Inset: Power spectral density of the mechanical motion for the green (drive power = 0.54 mW) and orange (drive power = 0.68 mW) markers, respectively. (b) Phase coherent oscillations are visible within the time-domain signal; here, the drive power is set above the parametric instability threshold. (c) With the drive power set below the parametric instability threshold, the time-domain signal comprises primarily of phase-incoherent mechanical noise. 
}
\label{Fig:07}
\end{figure}

We can consider the effect magnon backaction has on the mechanical decay rate. As described above, the interaction with magnons causes additional damping of the mechanical mode, $\Gamma_{\textrm{mag}} = 2\textrm{Im}\Sigma[\omega]$, which results in an effective mechanical damping rate,
\begin{equation}
    \Gamma_{\textrm{Tot}} = \Gamma_{\textrm{b}} + \Gamma_{\textrm{mag}}.
    \label{Decay}
\end{equation}
\noindent Just as in cavity optomechanics, $\Gamma_{\textrm{mag}}$ can be positive or negative \cite{aspelmeyer2014cavity}, thus either increasing or decreasing the total mechanical damping rate. The backaction enhancement of damping will be discussed in the next section; here, we will focus on the case of anti-damping. 

We now consider the data presented in Fig.~\ref{Fig:04}(b); the drive tone is on resonance with the upper normal mode, and the splitting between the normal modes is tuned to exactly one mechanical frequency. The additional magnomechanical anti-damping is thus maximized due to the triple-resonance enhancement and the total mechanical damping should decrease linearly with drive power as predicted by Eq.~\eqref{SelfEnergy}. This behavior is confirmed by the experimental points shown in Fig.~\ref{Fig:07}(a). Furthermore, fits to the total linewidth allow extraction of the intrinsic linewidth, $\Gamma_{\textrm{b}}$, as well as $g_{\textrm{mb}}^0$. Extrapolating to zero drive power yields, $\Gamma_{\textrm{b}} / 2\pi = 247$ Hz in air, $\Gamma_{\textrm{b}} / 2\pi = 98$ Hz in a partial pressure of helium ($\sim$ 15 Torr), and $\Gamma_{\textrm{b}} / 2\pi = 59$ Hz in vacuum; suggesting the primary damping mechanism was viscous air damping. The magnomechanical coupling rate can be determined using Eq.~\eqref{SelfEnergy} and \eqref{Decay} as well as the slope of the data presented in Fig.~\ref{Fig:07}(a) resulting in a value $g_{\textrm{mb}}^0 / 2\pi = 4.58$ mHz. As expected, the magnon-phonon coupling rate is independent of the intrinsic decay rate and is in good agreement with our numerical prediction and the result from the MMIT measurement (Appendix~\ref{Transparency}).

As the drive power is increased, a threshold will be reached where $\Gamma_{\textrm{b}} + \Gamma_{\textrm{mag}}$ becomes negative. In this situation, the mechanical oscillations will grow exponentially in time and will ultimately be limited by higher-order nonlinear effects. This parametric instability is analogous to lasing and is often referred to as phonon lasing \cite{ding2019phonon,mahboob2013phonon,kepesidis2013phonon,kemiktarak2014mode}. The onset of lasing can clearly be seen in Fig.~\ref{Fig:07}(a) as the total decay rate approaches zero above a threshold drive power. Furthermore, the inset of Fig.~\ref{Fig:07}(a) shows the mechanical power spectrum above (orange) and below (green) the threshold power. The onset of mechanical lasing results in four orders of magnitude increase of the mechanical power spectrum. The additional noise peaks in the lasing spectrum are a result of 60 Hz line noise captured by the solenoid being transduced via the mechanical mode. Finally, we observe the onset of mechanical lasing directly in the time-domain. The time-domain signal captured by the ADC is plotted in Fig.~\ref{Fig:07}(b,c). Below the lasing threshold, the signal is mainly comprised of noise; however, above lasing threshold, coherent oscillations at the mechanical frequency are visible \cite{spencer1958magneto,wang1970spin}. The time-domain and power spectrum data provide unambiguous evidence of phonon lasing, which could be used for stable clock signals \cite{vahala2009phonon}, or as the basis of sensitive mass and force sensors \cite{braginskiui1988resolution}.

\subsection{\label{Squash}Magnomechanical Cooling}

\begin{figure}[b]
\includegraphics[width = 0.48\textwidth]{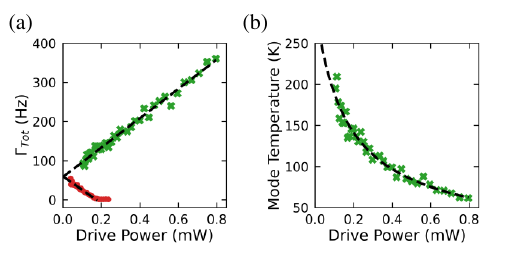}
\caption{Magnomechanical cooling. (a) Mechanical linewidth as a function of drive power in vacuum. Red circles demonstrate magnomechanical anti-damping and includes the data from Fig.~\ref{Fig:04}(b), green crosses demonstrate magnomechanical damping and includes the data from Fig.~\ref{Fig:04}(c). (b) Effective mode temperature of the mechanical mode determined from Eq.~\eqref{Temp}.}
\label{Fig:08}
\end{figure}

When driving on the red-sideband in our system, for example, Fig.~\ref{Fig:04}(c,e), the mechanical spectrum dips below the measurement noise floor, a phenomenon known as noise squashing \cite{aspelmeyer2014cavity}. Noise squashing has been observed in optomechanics, primarily in the context of feedback cooling \cite{bushev2006feedback,poggio2007feedback}. In feedback cooling, noise squashing results from the detector noise and the noise-driven mechanical motion becoming correlated. 

In our experiment, the detector noise was not fed into the system and cannot correlate with the mechanical motion. There was no feedback, and as a result, the noise squashing we have observed has a different origin. Indeed, it has a backaction-cooling origin, which has been observed in a microwave optomechanical nanobeam device \cite{rocheleau2010preparation}; however, it has not been observed in magnomechanics. Notably, because our experiment was performed at room temperature, there exists a large number of thermally excited gigahertz magnons and photons, $\bar{a}_{\rm{th}} \approx \bar{m}_{\rm{th}} \approx 800$. This thermal population produces a broad peak in the power spectrum, corresponding to the hybrid system's normal modes. In the triply resonant situation, the mechanical mode lies directly in the center of this broad peak in the power spectrum. Noise squashing results from destructive interference between upconverted drive excitations and the thermal excitations, causing the mechanical peak to appear below the detection noise floor. However, since the thermal peak is approximately constant over the mechanical mode's width, it is possible to extract the mechanical linewidth by performing a fit to an inverted power spectral density. The extracted linewidth from the data in Fig.~\ref{Fig:04}(b,c) are shown in Fig.~\ref{Fig:08}(a). Extrapolating to zero drive power, the intrinsic linewidth in vacuum from the damping and anti-damping data were $\Gamma_{\textrm{b}} / 2\pi = 58.5$ Hz and $\Gamma_{\textrm{b}} / 2\pi = 59.7$ Hz, respectively, which are in excellent agreement.

Finally, since our experiment lies well within the sideband resolved regime (i.e.~$\{\kappa,\gamma_m\} \ll \Omega_b$), and the number of thermal excitations is small compared to the phonon population, we can extract the effective mode temperature due to magnomechanical cooling. 
The effective phonon mode temperature is given by an expression similar to that for driven cavity optomechanical systems
\cite{aspelmeyer2014cavity}:
\begin{equation}
    T_{\textrm{final}} = T_{\textrm{init}} \bigg( \frac{\Gamma_{\textrm{b}}(1+\beta(\mathcal{P}))}{\Gamma_{\textrm{Tot}}} \bigg).
    \label{Temp}
\end{equation}
\noindent Here, $\beta(\mathcal{P})$ is a drive dependant variable; for derivation see Appendix~\ref{EffectiveTemp}. The effective mode temperature is defined in a consistent way for a driven system via the power spectrum of the mode \cite{clerk2010introductionto}. The effective mode temperature is shown in Fig.~\ref{Fig:08}(b); at the highest drive power the mechanical mode was cooled to approximately 65 kelvin from room temperature. With improvements to the experimental setup, such as smaller YIG spheres, and pre-cooling the experiment via cryogenics, which improves $\kappa$, $\Gamma_b$, and reduces thermal noise, it may be possible to achieve ground-state cooling of the mechanical vibrations.  Furthermore, with our current experimental values Eq.~\eqref{F} predicts a triple-resonance cooperativity enhancement of $F \approx 100$. Thus at cryogenic temperatures a cooperativity of $C \geq 1000$ is expected, placing our triply-resonant system among the state-of-the-art electromechanical experiments that recently have demonstrated macroscopic entanglement \cite{kotler2021direct}.

\section{Conclusion}

The magnomechanical interaction has in recent years been the focus of considerable theoretical work, yet experimental progress has been surprisingly limited. Here, we demonstrated the direct detection of mechanical vibration within a sub-mm YIG sphere and explored the full suite of dynamical backaction effects. We have shown that the magnomechanical interaction can amplify and cool the mechanical vibrations of the material effectively. As a consequence, we have observed phenomena such as microwave-driven phonon lasing, and noise squashing due to correlations with thermal noise. Unlike previous investigations, our experiment eases the observation of such dynamical backaction effects due both to the detection scheme and the specific configuration of the setup, in which the magnetic element is free to move in vacuum. These improvements have allowed the detection of magnon-induced mechanical frequency shifts, i.e.~the magnon spring effect. The experimental results agree well with our previous theoretical description of cavity magnomechanics and highlight the potential of cavity magnomechanical systems. With further improvement, microwave-driven magnomechanical cooling may be used to reach effective temperatures low enough to reach the ground state of the mechanics. This would not only allow the observation of quantum effects in a relatively massive system, but have also applications in quantum technologies such as quantum memories.

\begin{acknowledgments}
Authors acknowledge helpful contributions from D. Lachance-Quirion, D. Milling, and C. Doolin. This work was supported by the University of Alberta; the Natural Sciences and Engineering Research Council, Canada (Grants No.~RGPIN-04523-16 and No.~CREATE-495446-17); and the Alberta Quantum Major Innovation Fund. V.A.S.V. Bittencourt and S. Viola Kusminskiy acknowledge financial support from the Max Planck Society and from the Deutsche Forschungsgemeinschaft (DFG, German Research Foundation) through Project-ID 429529648–TRR 306 QuCoLiMa (“Quantum Cooperativity of Light and Matter”).
\end{acknowledgments}

\appendix

\section{\label{Transparency}Magnomechanically induced transparency}

\begin{figure}[t]
\includegraphics[width = 0.48\textwidth]{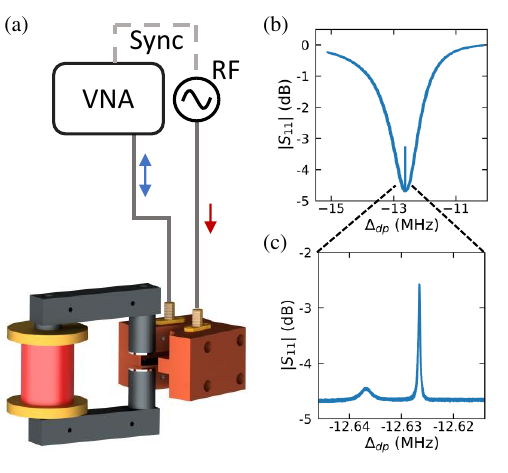}
\caption{Magnomechanically induced transparency. (a) Schematic illustration of the measurement setup: VNA, vector network analyzer; RF, microwave generator. The VNA and source are synced using a 10 MHz clock signal. Included is a rendering of the microwave cavity and magnetic yoke assembly. The frequency axis is negative due to the relative detuning between the drive tone and the normal mode. (b) Normalized reflection spectrum centred around the lower-normal mode, the system is tuned near spectrum (c) from Fig.~\ref{Fig:02}, but at a larger solenoid current to isolate the `magnon-like' normal mode. A narrow transparency window opens as $\Delta_{\textrm{dp}} = - \Omega_{\textrm{b}}$ due to the magnomechanical coupling. (c) Zoom-in of (b) showing a detailed spectrum of the magnomechanical induced transparency.}
\label{Fig:03}
\end{figure}

Since the mechanical mode probed in our experiment is different than in Ref.~\cite{zhang2016cavity}, to determine the mechanical frequency and extract the magnomechanical coupling rate, we calibrate the system using MMIT. MMIT is analogous to optomechanically induced transparency (OMIT) \cite{weis2010optomechanically,safavi2011electromagnetically}, and is a consequence of the interference of sidebands generated by the parametric coupling to phonons. Besides the natural response of the system at the normal mode frequencies, each mode has sidebands shifted by the phonon frequency. MMIT is observed by driving the cavity resonant with the red-sideband of one of the normal modes while sweeping the probe through the normal mode resonance. The interference between the weak probe and the upconverted excitations via the annihilation of a phonon generates a transparency window.

Figure~\ref{Fig:03}(a) illustrates a schematic of our MMIT measurement apparatus. A two-port microwave cavity was used, with the microwave drive at frequency $\omega_{\textrm{d}}$ connected to coupling port one and driven with a power between 1 and 50 mW. The vector network analyzer (VNA) probe, at frequency $\omega_{\textrm{p}}$, was connected to coupling port two, and the probe tone was held at a constant power of 0.03 mW.

To simplify our analysis, we decided to apply a red-detuned drive tone on the lower normal mode. The reflection spectrum as a function of the two-photon detuning $\Delta_{\textrm{dp}} = \omega_{\textrm{d}} - \omega_{\textrm{p}}$ is shown in Fig.~\ref{Fig:03}(b,c). A sharp peak can be seen at  $\Delta_{\textrm{dp}} = -\Omega_{\textrm{b}}$ resulting from the coherent magnomechanical interaction. A series of data was taken at atmospheric pressure for various pump powers and magnon-photon detunings, and fit using the theory presented in Ref.~\cite{zhang2016cavity}; described in Appendix~\ref{MMIT_Fit}. From this we were able to extract the single magnon-phonon coupling rate $g_{\textrm{mb}}^0 / 2\pi = 4.38$ mHz, the mechanical frequency $\Omega_{\textrm{b}} = 12.6270$ MHz, and the intrinsic mechanical decay rate $\Gamma_{\textrm{b}}/2\pi = 286$ Hz.

Fig.~\ref{Fig:03}(c) shows that there exists a second, higher frequency mechanical mode. This mode was not observed in the homodyne mechanical detection scheme, described in the main text, except at the highest drive powers. However, at those drive powers magnon nonlinearities resulted in the system becoming bistable and are therefore not included in our analysis \cite{wang2018bistability}. The second mechanical mode has a frequency  $\Omega_{\textrm{b}} = 12.637$ MHz, and a coupling rate $g_{\textrm{mb}}^0 / 2\pi = 2.41$ mHz. The difference in the transparency window height can be attributed to the increased damping rate, which is approximately an order of magnitude larger than the lower-frequency mode.  Numerical simulations reveal that clamping causes the $\textrm{S}_{1,2,0}$ mode to split into two nearly degenerate modes, resulting in the two modes observed here.

\section{\label{MMIT_Fit}{MMIT Theory}}

Here, we will outline the equations used to fit the magnomechanically induced transparency data; however, for a full description of this theory see the supplementary material included with Ref.~\cite{zhang2016cavity}. For the data presented in Section~\ref{Transparency}, the drive was detuned from the lower normal mode by the phonon frequency, $\omega_{\rm{d}} = \omega_- - \Omega_{\rm{b}}$. In this scenario the transparency window will have a peak reflectivity defined as
\begin{equation}
    r = \frac{1 - 2\frac{\kappa_{-,\rm{e}}}{\kappa_-}+C}{1+C}.
    \label{reflection}
\end{equation}
The transparency window can be seen in Fig.~\ref{Fig:03}(b,c). Here $\kappa_-$ and $\kappa_{-,e}$ are the linewidth and external coupling rate of the lower normal mode, respectively, and $C$ is the cooperativity.

The cooperativity can be shown to have the form,

\begin{equation}
    C \approx \frac{4\mathcal{P}(g_{\rm{mb}}^0)^2}{\hbar \omega_{\rm{d}} \Omega_{\rm{b}}^2 \Gamma_{\rm{b}}} \frac{\kappa_1 \sin^4(\theta)\cos^2(\theta)}{\kappa\cos^2(\theta) + \gamma_{\rm{m}}\sin^2(\theta)},
    \label{coop}
\end{equation}
where $\mathcal{P}$ is the microwave power at the experimental device. All losses have been carefully calibrated to ensure the accurate determination of the power reaching the device from the microwave sources. All other variables have been defined in the main text except $\theta \in [0,\pi/2]$, which is defined as,
\begin{equation}
    \tan(2\theta) = \frac{2g_{\rm{am}}}{\omega_{\rm{m}} -\omega_{\rm{a}}},
\end{equation}
where $\theta$ describes the hybridization of the normal modes; for maximally hybridized modes $\theta = \pi/4$.

\begin{figure}[b]
\includegraphics[width = 0.45\textwidth]{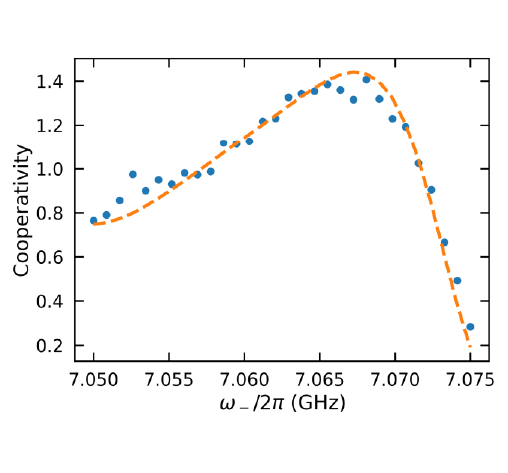}
\caption{ Magnomechanical cooperativity as a function of the lower normal mode frequency. For each measurement the drive tone is detuned by the phonon frequency from the lower normal mode, $\omega_{\rm{d}} = \omega_- - \Omega_{\rm{b}}$. Blue circles are experimentally determined using Eq.~\eqref{reflection}, and the dotted orange line is a numerical fit using Eq.~\eqref{coop}, where $g_{\rm{mb}}^0$ is the only fit parameter.}
\label{Fig:11}
\end{figure}

By varying the static magnetic field, and therefore the magnon-photon detuning, and measuring the cooperativity at each detuning using Eq.~\eqref{reflection} it is possible to determine the magnon-phonon coupling rate. The cooperativity as a function of the lower normal mode frequency is shown in Fig.~\ref{Fig:11}.

\section{\label{Heating}Magnon Heating}

\begin{figure}[b]
\includegraphics[width = 0.48\textwidth]{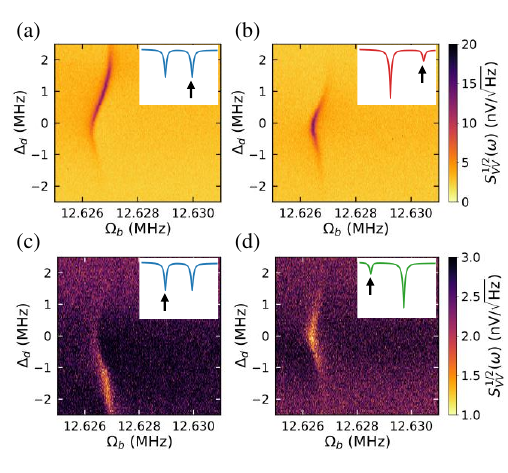}
\caption{Mechanical spectrum. (a-d) Mechanical power spectral density. Detuning is relative to the specific mode shown within the inset. In plots (a) and (c) the probe power is held constant at 0.54 mW. In plots (b) and (d) the probe power is held constant at 0.27 mW. All data presented was obtained with the experimental setup in atmospheric conditions.}
\label{Fig:05}
\end{figure}

In Fig.~\ref{Fig:05}, we observe a drive-dependent phonon frequency shift; however, this frequency is \textit{not} entirely due to dynamical magnon backaction. The observed frequency shift is mainly influenced by the heating of the YIG sample by the microwave drive. This is supported by the results plotted in Fig.~\ref{Fig:04}(b,c); for zero drive detuning in the triple-resonance scenario Eq.~\eqref{SelfEnergy} predicts zero frequency shift.  However, both cases result in a softening of the mechanical motion, likely due to a modification of the Young's modulus due to heating \cite{gibbons1958acoustical}. Magnetostrictive materials have been shown to exhibit a magnetic field dependant Young's modulus due to the $\Delta$E-effect \cite{honda1905change,scheidler2016dynamically}. Therefore, temperature-dependent modifications of the saturation magnetization and therefore internal static magnetic field may cause the observed frequency shift \cite{hansen1974saturation}. Furthermore, for all detunings presented in Fig.~\ref{Fig:05}, we observe a softening of the effective spring constant. However, dynamical backaction does not predict softening in all cases. Specifically, in Fig.~\ref{Fig:05}(c), we expect the magnon-spring effect to result in a hardening of the mechanics.

In all cases, the frequency shift follows the normal mode shape; as the normal mode depth increases, the circulating power increases, and the frequency shift increases. Moreover, the amplitude of the frequency shift is much larger than what is predicted by the calculated phonon self-energy in Eq.~\eqref{SelfEnergy} (see also \cite{potts2020magnon}). The maximum frequency shift predicted is approximately an order of magnitude smaller than the one observed. Thus, in this situation, any frequency shift resulting from the magnon-spring effect is overwhelmed by the additional frequency shift resulting from heating. 

The temperature dependence of the mechanical frequency was measured by heating the experimental setup. We observe a dependence of approximately $-715$ Hz/K for the frequency shift, suggesting that at the highest drive powers, the temperature of the sphere was increased by approximately 1 kelvin due to the microwave drive.  As described in the main text, this heating can be mitigated by using lower drive powers, placing the sphere in a partial pressure helium environment, and by applying the drive tone to the `photon-like' normal mode.  

Additionally, in Fig.~\ref{Fig:05} one may expect that the signal-to-noise ratio should be symmetric about zero detuning, with variations resulting from the homodyne detection sensitivity. However, the primary determining factor of the signal-to-noise ratio is related to the triple-resonance condition. For example, in Fig.~\ref{Fig:05}(a), and (c), the magnon-photon detuning was zero, such that the normal modes were fully hybridized, see Fig.~\ref{Fig:02}(b). In this situation, the normal mode spacing is slightly smaller than the phonon frequency $2g_{\textrm{am}} /2\pi = 10.86$ MHz and $\Omega_{\textrm{b}} / 2\pi = 12.627$ MHz. The asymmetry in the signal-to-noise ratio in Fig.~\ref{Fig:05}(a) can be understood by considering the mechanical sideband created on the microwave carrier. For negative drive detunings -- relative to the normal mode central frequency, i.e. $\Delta_{\textrm{d}} < 0$ -- the mechanical sideband is at a lower frequency than the lower normal mode and is therefore not resonantly enhanced. Conversely, a positive drive detuning ($\Delta_{\textrm{d}} > 0$) results in the lower mechanical sideband lying directly within the lower normal mode, resonantly enhancing this scatting process and improving the signal to noise ratio. A similar argument can be made for the scattering process regarding Fig.~\ref{Fig:05}(c), resulting in a resonant enhancement of the scattering process for negative detunings.

\section{\label{EffectiveTemp}Phonon temperature}

We now derive the effective phonon temperature and the approximations that lead to Eq.~\eqref{Temp}. The first step is to consider the linearized theory by writing the total Hamiltonian in terms of the fluctuations defined as $\hat{a} = \langle a \rangle + \delta \hat{a}$, $\hat{m} = \langle m \rangle + \delta \hat{m}$ and $\hat{b} = \langle b \rangle + \delta \hat{b}$, where the terms $\langle \cdot \rangle$ denote the steady-state average values. We then discard terms in the Hamiltonian containing more than two fluctuation terms and obtain the time domain Langevin equations of motion, which in the frequency domain read \cite{potts2020magnon}:
\begin{equation}
    \begin{aligned}
    \chi_a^{-1}[\omega] \delta \hat{a}[\omega] &= - i g_{\rm{am}} \delta \hat{m}[\omega] + \displaystyle \sum_{j = \rm{int},\, 1,\, 2} \sqrt{\kappa_j} \hat{\xi}_j[\omega] \\
    \chi_m^{-1}[\omega] \delta \hat{m}[\omega] &= - i g_{\rm{am}} \delta \hat{a}[\omega] - i g_{\rm{mb}} (\delta \hat{b}[\omega]+ \delta \hat{b}^\dagger[\omega]) \\
    &\quad + \sqrt{\gamma_m} \hat{\eta}[\omega] \\
    \chi_b^{-1}[\omega] \delta \hat{b}[\omega] &= - i (g_{\rm{mb}} \delta \hat{m}^\dagger [\omega] + g_{\rm{mb}}^* \delta \hat{m} [\omega]) \\ &+ \sqrt{\Gamma_b} \hat{\zeta}[\omega],
    \end{aligned}
\label{Lange}
\end{equation}
\noindent where the constants are defined within the main text. The total cavity decay rate $\kappa=\kappa_{\rm{int}} + \kappa_1 +\kappa_2$ is a sum of an intrinsic cavity decay $\kappa_{\rm{int}}$ and the decay rates into each of the ports $\kappa_{1,2}$. Accordingly, the total noise acting in the cavity is given by the terms $ \sqrt{\kappa_j} \hat{\xi}_j$, which includes the noises from each of the possible decay channels $\hat{\xi}_{\rm{int},1,2}$. Within our convention, the relation between time domain and frequency domain operators are $\hat{\mathcal{O}}(t) = \int d\omega e^{-i \omega t} \hat{\mathcal{O}}[\omega]$, and it should be noted that $\hat{\mathcal{O}}^\dagger[\omega] \equiv \int dt e^{i \omega t} \hat{\mathcal{O}}^\dagger (t) = (\hat{\mathcal{O}}[-\omega])^\dagger$.

For our analysis it is sufficient to consider a white thermal noise model described by the correlations \cite{clerk2010introductionto}:
\begin{equation}
\label{noise}
    \begin{aligned}
    \langle \hat{\xi}^\dagger_i[\omega^\prime] \hat{\xi}_i[\omega] \rangle &= 2 \pi n_{\rm{th}, a} \delta(\omega + \omega^\prime), \\
    \langle \hat{\xi}_i[\omega^\prime] \hat{\xi}^\dagger_i[\omega] \rangle &= 2 \pi (n_{\rm{th}, a} +1)\delta(\omega + \omega^\prime), \\
    \langle \hat{\eta}^\dagger[\omega^\prime] \hat{\eta}[\omega] \rangle &= 2 \pi n_{\rm{th}, m} \delta(\omega + \omega^\prime), \\
    \langle \hat{\eta}[\omega^\prime] \hat{\eta}^\dagger[\omega] \rangle &= 2 \pi (n_{\rm{th}, m}+1) \delta(\omega + \omega^\prime),\\
    \langle \hat{\zeta}^\dagger[\omega^\prime] \hat{\zeta}[\omega] \rangle &= 2 \pi n_{\rm{th}, b} \delta(\omega + \omega^\prime), \\
    \langle \hat{\zeta}[\omega^\prime] \hat{\zeta}^\dagger[\omega] \rangle &= 2 \pi (n_{\rm{th}, b}+1) \delta(\omega + \omega^\prime).
    \end{aligned}
\end{equation}
\noindent The thermal occupancy of the baths is given by Bose-Einstein distributions
\begin{equation}
\begin{aligned}
n_{{\rm{th}}, (a,m)} &= \frac{1}{{\rm{exp}}\left[\frac{\hbar \omega_{(a,m)}}{k_B T_{{\rm{Bath}}}}\right]-1}, \\
n_{{\rm{th}}, (b)} &= \frac{1}{{\rm{exp}}\left[\frac{\hbar \Omega_{b}}{k_B T_{{\rm{Bath}}}}\right]-1},
\end{aligned}
\end{equation}
\noindent where $k_B$ is the Boltzmann constant, $T_{{\rm{Bath}}}$ is the bath temperature (assumed to be the same for all the modes),  and we have assumed that the occupancy of the baths for all decay channels are the same and given in terms of the cavity frequency $\omega_{\rm{a}}$. This last assumption can be readily generalized.

Before deriving the effective temperature of the phonon mode in the coupled and driven system described by the Eq.~\eqref{Lange}, let us first consider why the noise spectral density of the phonon mode provides information about that mode's temperature. For that, we consider the simpler problem in which there is no magnon-phonon coupling. In this case the phonon mode is driven only by thermal noise, and the phonon component of Eq.~\eqref{Lange} reads
\begin{equation}
\begin{aligned}
\delta \hat{b}[\omega] &= \chi_b [\omega] \sqrt{\Gamma_b} \hat{\zeta}[\omega], \\
\delta \hat{b}^\dagger [\omega] &= \chi_b^* [-\omega] \sqrt{\Gamma_b} \hat{\zeta}^\dagger[\omega].
\end{aligned}
\end{equation}
\noindent We then can consider the spectral density given by
\begin{equation}
\begin{aligned}
    S_{\delta b^\dagger \delta b}[\omega] &= \int_{-\infty}^\infty dt e^{i \omega t} \langle \delta \hat{b}^\dagger(t) \hat{b}(0) \rangle, \\ 
    &= \int_{-\infty} ^\infty \frac{d \omega^\prime}{2 \pi} \langle \delta \hat{b}^\dagger [\omega] \delta \hat{b} [\omega^\prime] \rangle.
\end{aligned}
\end{equation}
\noindent For the simple uncoupled case, the solutions of the frequency domain equations combined with the noise correlations Eq.~\eqref{noise} yields the simple relation,
\begin{equation}
S_{\delta b^\dagger \delta b}[\omega] = \frac{\Gamma_b n_{\rm{th}, b}}{(\omega + \Omega_b)^2 + \frac{\Gamma_b^2}{4}},  
\end{equation}
\noindent which is given in terms of the bath occupancy $n_{\rm{th}, b}$ and thus the temperature of the mode. The noise spectral density $S_{\delta b^\dagger \delta b}[-\Omega_b]$ and its counterpart $S_{\delta b \delta b^\dagger}[\Omega_b]$ are linked to the ability of the oscillator to emit/absorb energy \cite{marquardt2007quantumtheory,clerk2010introductionto}. 

Turning our attention to the full problem, we solve the linear Langevin equations \eqref{Lange} and obtain for the phonon mode
\begin{equation}
\label{fullsolv}
    \begin{aligned}
    \left(\chi_b^{-1}[\omega] - i \Sigma[\omega] \right)\delta \hat{b}[\omega] &= \mathcal{A}[\omega] \sum_i \sqrt{\kappa_i} \hat{\xi}_i [\omega] \\
    &\,\mkern-100mu+ \tilde{\mathcal{A}}[\omega] \sum_i \sqrt{\kappa_i} \hat{\xi}^\dagger [\omega] \\
    &\,\mkern-100mu + \mathcal{B}[\omega]\sqrt{\gamma_m} \hat{\eta}[\omega] + \tilde{\mathcal{B}}[\omega]\sqrt{\gamma_m} \hat{\eta}^\dagger[\omega] \\
    &\,\mkern-100mu +\sqrt{\Gamma_b} \hat{\zeta}[\omega] + \tilde{\mathcal{C}}[\omega]\sqrt{\Gamma_b} \hat{\zeta}^\dagger[\omega],
    \end{aligned}
\end{equation}
\noindent where $\Sigma[\omega]$ is the self energy given in the main text Eq.~\eqref{SelfEnergy} whose derivation is discussed in detail in Ref.~\cite{potts2020magnon}. The frequency-dependent coefficients have complicated and not elucidating forms that depend on the susceptibilities and on the couplings. We then rewrite the above expression as
\begin{equation}
\left(i (\tilde{\Omega}_b[\omega] -\omega) + \frac{\Gamma_{\rm{Tot}} [\omega]}{2} \right)\delta \hat{b}[\omega] = \sqrt{\Gamma_{\rm{Tot}} [\omega]}\hat{\Upsilon}[\omega],
\end{equation}
\noindent where
\begin{equation}
    \begin{aligned}
    \tilde{\Omega}_b[\omega] &= \Omega_b + \delta\Omega_b[\omega] = \Omega_b - {\rm{Re}} \Sigma[\omega], \\
    \Gamma_{\rm{Tot}}[\omega] &= \Gamma_b + \Gamma_{\rm{mag}}[\omega] = \Gamma_b + 2 {\rm{Im}}\Sigma[\omega],
    \end{aligned}
\end{equation}
\noindent are the phonon frequency and decay rates corrected by the self energy, and $\hat{\Upsilon}[\omega]$ is the combination of noises appearing in the right-hand side of Eq.~\eqref{fullsolv} divided by the total phonon decay rate.

\begin{figure}[b]
\includegraphics[width = 0.35\textwidth]{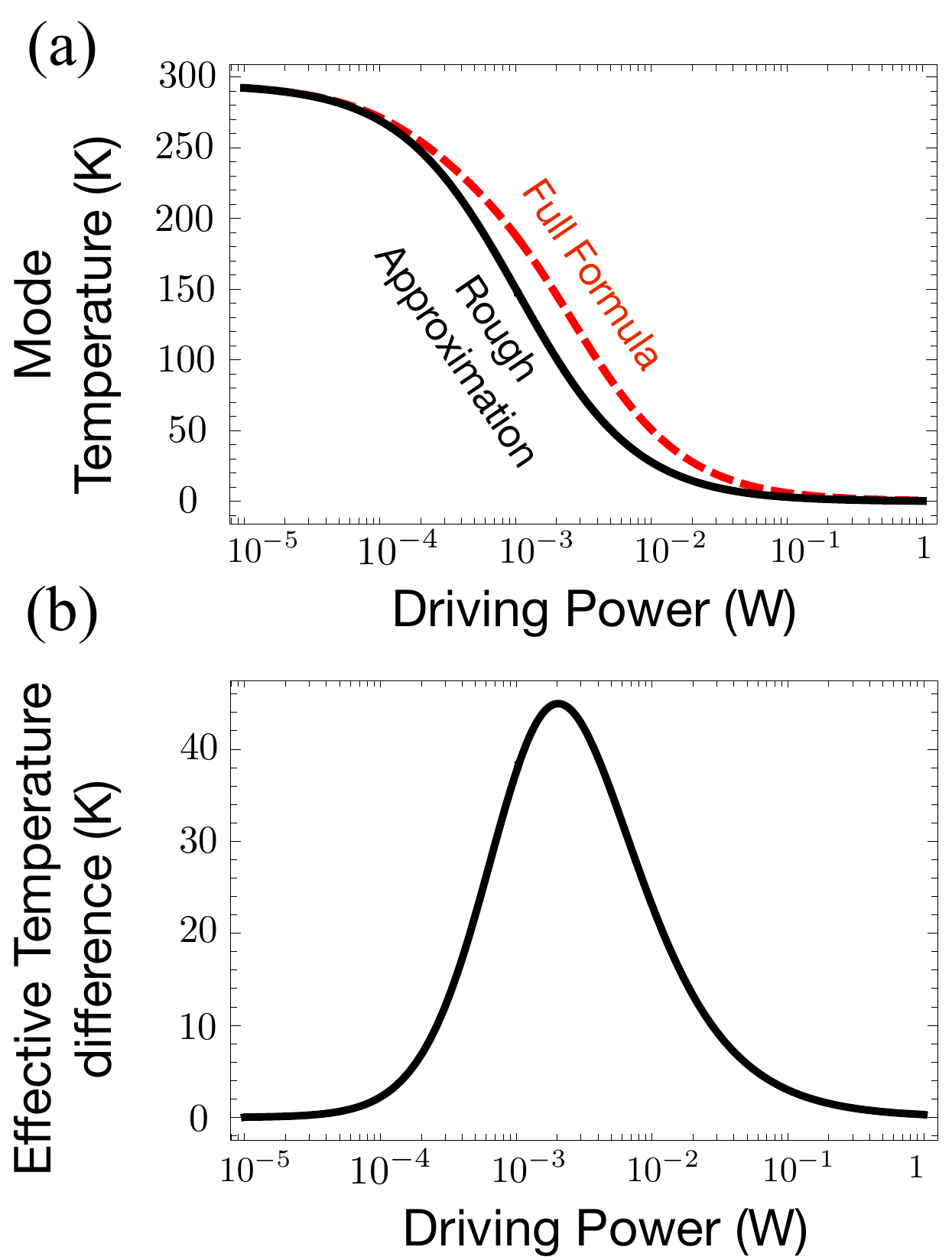}
\caption{Comparison between the full formula for the effective temperature Eq.~\eqref{effTemp} and the approximate estimate Eq.~\eqref{Rapprox}. (a) Effective temperature as given by equation Eq.~\eqref{effTemp} (dashed line) and as given by Eq.~\eqref{Rapprox} (continuous line) as a function of the driving power. (b) Difference between the effective temperatures (Eq.~\eqref{effTemp} - Eq.~\eqref{Rapprox}) as a function of the driving powers. For both plots we use the experimental parameters and driving scheme corresponding to the results of Fig~\ref{Fig:08}, and the bath temperature is $T_{\rm{Bath}} = 295$ K.}
\label{Fig:EffTempComp}
\end{figure}

Recalling the relation of the spectral density $S_{\delta b^\dagger \delta b}[\omega]$ to the thermal number of phonons and using the noise relations Eq.~\eqref{noise}, we obtain \cite{safavinaeini2013lasernoise}
\begin{equation}
S_{\delta b^\dagger \delta b}[\omega] = \frac{\Gamma_{\rm{Tot}}[\omega] n_{{\rm{eff}}, b}[\omega]}{(\omega + \tilde{\Omega}_b[\omega])^2 + \frac{\Gamma_{\rm{Tot}}^2[\omega]}{4}},  
\end{equation}
\noindent where the effective phonon number is given in terms of the frequency dependent coefficients of Eq.~\eqref{fullsolv} and of the thermal occupancy of the baths as
\begin{equation}
\label{FullN}
\begin{aligned}
    \Gamma_{\rm{Tot}}[\omega] n_{{\rm{eff}},b}[\omega] &= \kappa \vert \mathcal{A}[-\omega] \vert^2 n_{{\rm{th}}, a} \\
    &\, \mkern-50mu+ \kappa \vert \tilde{\mathcal{A}}[-\omega] \vert^2 (n_{{\rm{th}}, a}+1) \\
    &\,\mkern-50mu +\gamma_m \vert \mathcal{B}[-\omega] \vert^2 n_{{\rm{th}}, m} \\
    &\,\mkern-50mu + \gamma_m \vert \tilde{\mathcal{B}}[-\omega] \vert^2 (n_{{\rm{th}}, m}+1) \\
    &\,\mkern-50mu +\Gamma_b  n_{{\rm{th}}, b} + \Gamma_b \vert \tilde{\mathcal{C}}[-\omega] \vert^2 (n_{{\rm{th}}, b}+1).
\end{aligned}
\end{equation}
\noindent In deriving the above formula we have assumed that all the decay channels of the cavity mode are related to thermal baths at the same temperature. The effective temperature for the phonon mode is then given by
\begin{equation}
\label{effTemp}
T_{{\rm{eff}},b}[\omega] = \frac{\hbar \tilde{\Omega}_b[\omega]}{k_B} \left[{\rm{ln}} \left(\frac{n_{{\rm{eff}},b}[\omega]+1}{n_{{\rm{eff}},b}[\omega]} \right) \right]^{-1}.
\end{equation}
\noindent We can make further approximations to the effective temperature formula. Since $\omega_{(a,m)}$ are three orders of magnitude larger than the phonon frequency, $n_{{\rm{th}}, (a,m)} \ll n_{{\rm{th}}, b}$ and we can discard the terms $\propto n_{{\rm{th}}, (a,m)}$ and $\propto n_{{\rm{th}}, (a,m)} +1$. Furthermore, at room temperature $n_{{\rm{th}}, b} \gg 1$, and Eq.~\eqref{FullN} simplifies to
\begin{equation}
\Gamma_{\rm{Tot}}[\omega] n_{{\rm{eff}},b}[\omega] = \Gamma_b (1 + \vert \tilde{\mathcal{C}}[-\omega] \vert^2) n_{{\rm{th}},b}.
\end{equation}
\noindent The remaining frequency-dependent coefficient is given explicitly by
\begin{equation}
\label{consts01}
\begin{aligned}
\tilde{\mathcal{C}}[\omega]&= \frac{\vert g_{{\rm{mb}}}\vert^2 \chi_b^*[-\omega] \left(\Xi^*[-\omega] - \Xi[\omega]\right)}{1+\vert g_{{\rm{mb}}}\vert^2 \chi_b^*[-\omega]\left(\Xi^*[-\omega] - \Xi[\omega]\right)},
\end{aligned}
\end{equation}
\noindent where $\Xi[\omega] = [\chi^{-1}_{\textrm{m}}[\omega] + g^2_{\textrm{am}}\chi_{\textrm{a}}[\omega]]^{-1}$, and Eq.~\eqref{consts01} depends on the driving power only through $\vert g_{{\rm{mb}}}\vert^2 = \vert g_{\rm{mb}}^0  \langle m \rangle \vert^2 $ where \cite{potts2020magnon}:
\begin{equation}
\label{consts02}
\langle m \rangle = \frac{i \epsilon_{\rm{d}} \sqrt{\kappa_{\rm{ext}}}}{(i \Delta_{\rm{a}} - \kappa/2)(i \Delta_{\rm{m}} - \gamma_m/2) + g_{\rm{am}}^2}.
\end{equation}

At low powers, the contribution $\propto \vert \tilde{\mathcal{C}}[-\omega] \vert^2$ can be safely discarded, but as the power increases, since $\chi_b[\omega]$ is sharply peaked around $\Omega_b$, this contribution becomes prominent. In fact, $\vert \tilde{\mathcal{C}}[-\omega] \vert^2$ goes from zero to its maximum value of one. In the limit $k_B T \gg \hbar \omega_{a,b,m}$, valid for our room temperature experiment, we can write
\begin{equation}
\begin{aligned}
n_{{\rm{eff}},b}[\omega] &\sim \frac{k_B T_{{\rm{eff}},b}[\omega] }{\hbar \tilde{\Omega}_b[\omega]}, \\
n_{{\rm{th}}, (b)} &\sim \frac{k_B T_{{\rm{Bath}}}}{\hbar \Omega_b},
\end{aligned}
\end{equation}
\noindent and since the phonon frequency shift $\delta \Omega_b \ll \Omega_b$, we can take $\tilde{\Omega}_b \sim \Omega_b$. Within those approximations, the phonon mode effective temperature reads
\begin{equation}
\label{improv}
T_{{\rm{eff}},b}[\omega] \approx \frac{\Gamma_b T_{{\rm{Bath}}} (1 + \vert \tilde{\mathcal{C}}[-\omega] \vert^2)}{\Gamma_{\rm{Tot}}[\omega]}.
\end{equation}
\begin{figure}[b]
\includegraphics[width = 0.35\textwidth]{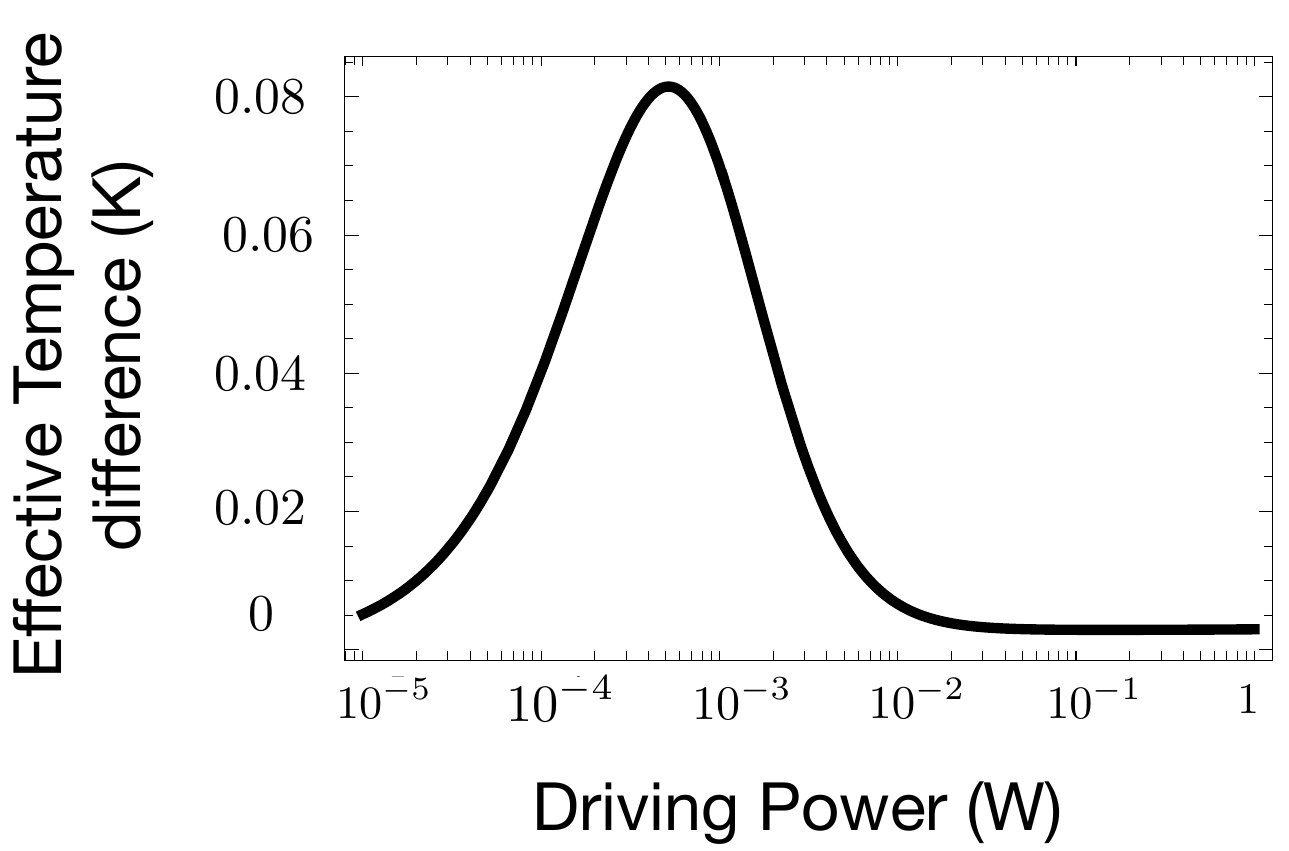}
\caption{Difference between the effective temperature given by the full formula Eq.~\eqref{effTemp} and the improved approximation Eq.~\eqref{improv}. For this plot we have used the experimental parameters and driving scheme corresponding to the results of Fig.~\ref{Fig:08}, and the bath temperature is $T_{\rm{Bath}} = 295$ K.}
\label{Fig:EffTempComp2}
\end{figure}
\noindent A rough estimate of the effective temperature can be made by discarding the term $\propto \vert \tilde{\mathcal{C}}[-\omega] \vert^2$, such that
\begin{equation}
\label{Rapprox}
T_{{\rm{eff}},b}[\omega] \approx \frac{\Gamma_b T_{{\rm{Bath}}} }{\Gamma_{\rm{Tot}}[\omega]}.
\end{equation}
\noindent This has a familiar form of the effective phonon temperature in driven optomechanical systems (c.f. \cite{aspelmeyer2014cavity,safavinaeini2013lasernoise}), and in our case it is valid for low driving powers only. This can be seen in Fig.~\ref{Fig:EffTempComp} which shows the effective temperature at the phonon frequency for the red detuning scheme (driving at the lower normal mode) as given by Eq.~\eqref{effTemp} and as given by Eq.~\eqref{Rapprox} for the parameters values corresponding to Fig.~\ref{Fig:08}. The approximation is good for small driving powers but at powers larger than 0.1 mW, the difference between the full formula and the rough approximation can be $\sim 40$ K. Thus, even though Eq.~\eqref{Rapprox} is a simple and practical approximation, it leads to an underestimate of the phonon effective temperature.

The improved approximation Eq.~\eqref{improv} adds the contribution $\propto \vert \tilde{\mathcal{C}}[-\omega] \vert^2$ which, for a given detuning and at $\Omega_b$ is a function of the driving power $\mathcal{P}$ given by
\begin{equation}
    \vert \tilde{\mathcal{C}}[-\Omega_b] \vert^2= \frac{\mathcal{P}^2 \vert \mathbb{C} \vert^2}{1 + 2 \mathcal{P} {\rm{Re}}\left[ \mathbb{C} \right]+ \mathcal{P}^2 \vert \mathbb{C} \vert^2},
\end{equation}
\noindent where $\mathbb{C}$ is a complex number that can be calculated with Eq.~\eqref{consts01} and Eq.~\eqref{consts02} for a given set of parameters. For conciseness, in the main text we define $\vert \tilde{\mathcal{C}}[-\Omega_b] \vert^2$ as $\beta(\mathcal{P})$. Figure \ref{Fig:EffTempComp2} shows the difference between the effective temperature obtained by the full formula Eq.~\eqref{effTemp}, and the approximation including the aforementioned power dependent factor Eq.~\eqref{improv}. In this case, the maximum difference is now $\sim 0.08$ K, for driving powers corresponding to temperatures of $\sim 150$ K, and thus this approximation is more suitable for analysing the experimental data.

\bibliography{apssamp}

\providecommand{\noopsort}[1]{}\providecommand{\singleletter}[1]{#1}%
\begin{thebibliography}{76}%
\makeatletter
\providecommand \@ifxundefined [1]{%
 \@ifx{#1\undefined}
}%
\providecommand \@ifnum [1]{%
 \ifnum #1\expandafter \@firstoftwo
 \else \expandafter \@secondoftwo
 \fi
}%
\providecommand \@ifx [1]{%
 \ifx #1\expandafter \@firstoftwo
 \else \expandafter \@secondoftwo
 \fi
}%
\providecommand \natexlab [1]{#1}%
\providecommand \enquote  [1]{``#1''}%
\providecommand \bibnamefont  [1]{#1}%
\providecommand \bibfnamefont [1]{#1}%
\providecommand \citenamefont [1]{#1}%
\providecommand \href@noop [0]{\@secondoftwo}%
\providecommand \href [0]{\begingroup \@sanitize@url \@href}%
\providecommand \@href[1]{\@@startlink{#1}\@@href}%
\providecommand \@@href[1]{\endgroup#1\@@endlink}%
\providecommand \@sanitize@url [0]{\catcode `\\12\catcode `\$12\catcode
  `\&12\catcode `\#12\catcode `\^12\catcode `\_12\catcode `\%12\relax}%
\providecommand \@@startlink[1]{}%
\providecommand \@@endlink[0]{}%
\providecommand \url  [0]{\begingroup\@sanitize@url \@url }%
\providecommand \@url [1]{\endgroup\@href {#1}{\urlprefix }}%
\providecommand \urlprefix  [0]{URL }%
\providecommand \Eprint [0]{\href }%
\providecommand \doibase [0]{https://doi.org/}%
\providecommand \selectlanguage [0]{\@gobble}%
\providecommand \bibinfo  [0]{\@secondoftwo}%
\providecommand \bibfield  [0]{\@secondoftwo}%
\providecommand \translation [1]{[#1]}%
\providecommand \BibitemOpen [0]{}%
\providecommand \bibitemStop [0]{}%
\providecommand \bibitemNoStop [0]{.\EOS\space}%
\providecommand \EOS [0]{\spacefactor3000\relax}%
\providecommand \BibitemShut  [1]{\csname bibitem#1\endcsname}%
\let\auto@bib@innerbib\@empty
\bibitem [{\citenamefont {Aspelmeyer}\ \emph {et~al.}(2014)\citenamefont
  {Aspelmeyer}, \citenamefont {Kippenberg},\ and\ \citenamefont
  {Marquardt}}]{aspelmeyer2014cavity}%
  \BibitemOpen
  \bibfield  {author} {\bibinfo {author} {\bibfnamefont {M.}~\bibnamefont
  {Aspelmeyer}}, \bibinfo {author} {\bibfnamefont {T.~J.}\ \bibnamefont
  {Kippenberg}},\ and\ \bibinfo {author} {\bibfnamefont {F.}~\bibnamefont
  {Marquardt}},\ }\bibfield  {title} {\bibinfo {title} {\textit{Cavity
  optomechanics}},\ }\href {https://doi.org/10.1103/RevModPhys.86.1391}
  {\bibfield  {journal} {\bibinfo  {journal} {Rev. Mod. Phys.}\ }\textbf
  {\bibinfo {volume} {86}},\ \bibinfo {pages} {1391} (\bibinfo {year}
  {2014})}\BibitemShut {NoStop}%
\bibitem [{\citenamefont {Lachance-Quirion}\ \emph {et~al.}(2019)\citenamefont
  {Lachance-Quirion}, \citenamefont {Tabuchi}, \citenamefont {Gloppe},
  \citenamefont {Usami},\ and\ \citenamefont {Nakamura}}]{lachance2019hybrid}%
  \BibitemOpen
  \bibfield  {author} {\bibinfo {author} {\bibfnamefont {D.}~\bibnamefont
  {Lachance-Quirion}}, \bibinfo {author} {\bibfnamefont {Y.}~\bibnamefont
  {Tabuchi}}, \bibinfo {author} {\bibfnamefont {A.}~\bibnamefont {Gloppe}},
  \bibinfo {author} {\bibfnamefont {K.}~\bibnamefont {Usami}},\ and\ \bibinfo
  {author} {\bibfnamefont {Y.}~\bibnamefont {Nakamura}},\ }\bibfield  {title}
  {\bibinfo {title} {\textit{Hybrid quantum systems based on magnonics}},\
  }\href {https://doi.org/10.7567/1882-0786/ab248d} {\bibfield  {journal}
  {\bibinfo  {journal} {Appl. Phys. Express}\ }\textbf {\bibinfo {volume}
  {12}},\ \bibinfo {pages} {070101} (\bibinfo {year} {2019})}\BibitemShut
  {NoStop}%
\bibitem [{\citenamefont {Awschalom}\ \emph {et~al.}(2021)\citenamefont
  {Awschalom}, \citenamefont {Du}, \citenamefont {He}, \citenamefont
  {Heremans}, \citenamefont {Hoffmann}, \citenamefont {Hou}, \citenamefont
  {Kurebayashi}, \citenamefont {Li}, \citenamefont {Liu}, \citenamefont
  {Novosad} \emph {et~al.}}]{awschalom2021quantum}%
  \BibitemOpen
  \bibfield  {author} {\bibinfo {author} {\bibfnamefont {D.~D.}\ \bibnamefont
  {Awschalom}}, \bibinfo {author} {\bibfnamefont {C.}~\bibnamefont {Du}},
  \bibinfo {author} {\bibfnamefont {R.}~\bibnamefont {He}}, \bibinfo {author}
  {\bibfnamefont {F.}~\bibnamefont {Heremans}}, \bibinfo {author}
  {\bibfnamefont {A.}~\bibnamefont {Hoffmann}}, \bibinfo {author}
  {\bibfnamefont {J.}~\bibnamefont {Hou}}, \bibinfo {author} {\bibfnamefont
  {H.}~\bibnamefont {Kurebayashi}}, \bibinfo {author} {\bibfnamefont
  {Y.}~\bibnamefont {Li}}, \bibinfo {author} {\bibfnamefont {L.}~\bibnamefont
  {Liu}}, \bibinfo {author} {\bibfnamefont {V.}~\bibnamefont {Novosad}}, \emph
  {et~al.},\ }\bibfield  {title} {\bibinfo {title} {\textit{Quantum Engineering
  With Hybrid Magnonics Systems and Materials}},\ }\href
  {https://doi.org/10.1109/TQE.2021.3057799} {\bibfield  {journal} {\bibinfo
  {journal} {IEEE T. Quantum Eng.}\ ,\ \bibinfo {pages} {1}} (\bibinfo {year}
  {2021})}\BibitemShut {NoStop}%
\bibitem [{\citenamefont {Li}\ \emph {et~al.}(2020{\natexlab{a}})\citenamefont
  {Li}, \citenamefont {Zhang}, \citenamefont {Tyberkevych}, \citenamefont
  {Kwok}, \citenamefont {Hoffmann},\ and\ \citenamefont
  {Novosad}}]{li2020hybrid}%
  \BibitemOpen
  \bibfield  {author} {\bibinfo {author} {\bibfnamefont {Y.}~\bibnamefont
  {Li}}, \bibinfo {author} {\bibfnamefont {W.}~\bibnamefont {Zhang}}, \bibinfo
  {author} {\bibfnamefont {V.}~\bibnamefont {Tyberkevych}}, \bibinfo {author}
  {\bibfnamefont {W.-K.}\ \bibnamefont {Kwok}}, \bibinfo {author}
  {\bibfnamefont {A.}~\bibnamefont {Hoffmann}},\ and\ \bibinfo {author}
  {\bibfnamefont {V.}~\bibnamefont {Novosad}},\ }\bibfield  {title} {\bibinfo
  {title} {\textit{Hybrid magnonics: Physics, circuits, and applications for
  coherent information processing}},\ }\href
  {https://doi.org/10.1063/5.0020277} {\bibfield  {journal} {\bibinfo
  {journal} {J. Appl. Phys.}\ }\textbf {\bibinfo {volume} {128}},\ \bibinfo
  {pages} {130902} (\bibinfo {year} {2020}{\natexlab{a}})}\BibitemShut
  {NoStop}%
\bibitem [{\citenamefont {Ebrahimi}\ \emph {et~al.}(2020)\citenamefont
  {Ebrahimi}, \citenamefont {Motazedifard},\ and\ \citenamefont
  {Bagheri}}]{ebrahimi2020ultra}%
  \BibitemOpen
  \bibfield  {author} {\bibinfo {author} {\bibfnamefont {M.~S.}\ \bibnamefont
  {Ebrahimi}}, \bibinfo {author} {\bibfnamefont {A.}~\bibnamefont
  {Motazedifard}},\ and\ \bibinfo {author} {\bibfnamefont {M.}~\bibnamefont
  {Bagheri}},\ }\bibfield  {title} {\bibinfo {title} {\textit{Ultra-precision
  single-quadrature quantum magnetometry in cavity electromagnonics}},\ }\href
  {arXiv:2011.06081} {\bibfield  {journal} {\bibinfo  {journal}
  {arXiv:2011.06081}\ } (\bibinfo {year} {2020})}\BibitemShut {NoStop}%
\bibitem [{\citenamefont {Crescini}\ \emph {et~al.}(2020)\citenamefont
  {Crescini}, \citenamefont {Alesini}, \citenamefont {Braggio}, \citenamefont
  {Carugno}, \citenamefont {D’Agostino}, \citenamefont {Di~Gioacchino},
  \citenamefont {Falferi}, \citenamefont {Gambardella}, \citenamefont {Gatti},
  \citenamefont {Iannone} \emph {et~al.}}]{crescini2020axion}%
  \BibitemOpen
  \bibfield  {author} {\bibinfo {author} {\bibfnamefont {N.}~\bibnamefont
  {Crescini}}, \bibinfo {author} {\bibfnamefont {D.}~\bibnamefont {Alesini}},
  \bibinfo {author} {\bibfnamefont {C.}~\bibnamefont {Braggio}}, \bibinfo
  {author} {\bibfnamefont {G.}~\bibnamefont {Carugno}}, \bibinfo {author}
  {\bibfnamefont {D.}~\bibnamefont {D’Agostino}}, \bibinfo {author}
  {\bibfnamefont {D.}~\bibnamefont {Di~Gioacchino}}, \bibinfo {author}
  {\bibfnamefont {P.}~\bibnamefont {Falferi}}, \bibinfo {author} {\bibfnamefont
  {U.}~\bibnamefont {Gambardella}}, \bibinfo {author} {\bibfnamefont
  {C.}~\bibnamefont {Gatti}}, \bibinfo {author} {\bibfnamefont
  {G.}~\bibnamefont {Iannone}}, \emph {et~al.},\ }\bibfield  {title} {\bibinfo
  {title} {\textit{Axion search with a quantum-limited ferromagnetic
  haloscope}},\ }\href {https://doi.org/10.1103/PhysRevLett.124.171801}
  {\bibfield  {journal} {\bibinfo  {journal} {Phys. Rev. Lett.}\ }\textbf
  {\bibinfo {volume} {124}},\ \bibinfo {pages} {171801} (\bibinfo {year}
  {2020})}\BibitemShut {NoStop}%
\bibitem [{\citenamefont {Ikeda}\ \emph {et~al.}(2021)\citenamefont {Ikeda},
  \citenamefont {Ito}, \citenamefont {Miuchi}, \citenamefont {Soda},
  \citenamefont {Kurashige}, \citenamefont {Lachance-Quirion}, \citenamefont
  {Nakamura},\ and\ \citenamefont {Shikano}}]{ikeda2021axion}%
  \BibitemOpen
  \bibfield  {author} {\bibinfo {author} {\bibfnamefont {T.}~\bibnamefont
  {Ikeda}}, \bibinfo {author} {\bibfnamefont {A.}~\bibnamefont {Ito}}, \bibinfo
  {author} {\bibfnamefont {K.}~\bibnamefont {Miuchi}}, \bibinfo {author}
  {\bibfnamefont {J.}~\bibnamefont {Soda}}, \bibinfo {author} {\bibfnamefont
  {H.}~\bibnamefont {Kurashige}}, \bibinfo {author} {\bibfnamefont
  {D.}~\bibnamefont {Lachance-Quirion}}, \bibinfo {author} {\bibfnamefont
  {Y.}~\bibnamefont {Nakamura}},\ and\ \bibinfo {author} {\bibfnamefont
  {Y.}~\bibnamefont {Shikano}},\ }\bibfield  {title} {\bibinfo {title}
  {\textit{Axion search with quantum nondemolition detection of magnons}},\
  }\href@noop {} {\bibfield  {journal} {\bibinfo  {journal} {arXiv:2102.08764}\
  } (\bibinfo {year} {2021})}\BibitemShut {NoStop}%
\bibitem [{\citenamefont {Flower}\ \emph {et~al.}(2019)\citenamefont {Flower},
  \citenamefont {Bourhill}, \citenamefont {Goryachev},\ and\ \citenamefont
  {Tobar}}]{flower2019broadening}%
  \BibitemOpen
  \bibfield  {author} {\bibinfo {author} {\bibfnamefont {G.}~\bibnamefont
  {Flower}}, \bibinfo {author} {\bibfnamefont {J.}~\bibnamefont {Bourhill}},
  \bibinfo {author} {\bibfnamefont {M.}~\bibnamefont {Goryachev}},\ and\
  \bibinfo {author} {\bibfnamefont {M.~E.}\ \bibnamefont {Tobar}},\ }\bibfield
  {title} {\bibinfo {title} {\textit{Broadening frequency range of a
  ferromagnetic axion haloscope with strongly coupled cavity--magnon
  polaritons}},\ }\href
  {https://doi.org/https://doi.org/10.1016/j.dark.2019.100306} {\bibfield
  {journal} {\bibinfo  {journal} {Phys. Dark Universe}\ }\textbf {\bibinfo
  {volume} {25}},\ \bibinfo {pages} {100306} (\bibinfo {year}
  {2019})}\BibitemShut {NoStop}%
\bibitem [{\citenamefont {Huebl}\ \emph {et~al.}(2013)\citenamefont {Huebl},
  \citenamefont {Zollitsch}, \citenamefont {Lotze}, \citenamefont {Hocke},
  \citenamefont {Greifenstein}, \citenamefont {Marx}, \citenamefont {Gross},\
  and\ \citenamefont {Goennenwein}}]{huebl2013high}%
  \BibitemOpen
  \bibfield  {author} {\bibinfo {author} {\bibfnamefont {H.}~\bibnamefont
  {Huebl}}, \bibinfo {author} {\bibfnamefont {C.~W.}\ \bibnamefont
  {Zollitsch}}, \bibinfo {author} {\bibfnamefont {J.}~\bibnamefont {Lotze}},
  \bibinfo {author} {\bibfnamefont {F.}~\bibnamefont {Hocke}}, \bibinfo
  {author} {\bibfnamefont {M.}~\bibnamefont {Greifenstein}}, \bibinfo {author}
  {\bibfnamefont {A.}~\bibnamefont {Marx}}, \bibinfo {author} {\bibfnamefont
  {R.}~\bibnamefont {Gross}},\ and\ \bibinfo {author} {\bibfnamefont
  {S.~T.~B.}\ \bibnamefont {Goennenwein}},\ }\bibfield  {title} {\bibinfo
  {title} {\textit{High cooperativity in coupled microwave resonator
  ferrimagnetic insulator hybrids}},\ }\href
  {https://doi.org/10.1103/PhysRevLett.111.127003} {\bibfield  {journal}
  {\bibinfo  {journal} {Phys. Rev. Lett.}\ }\textbf {\bibinfo {volume} {111}},\
  \bibinfo {pages} {127003} (\bibinfo {year} {2013})}\BibitemShut {NoStop}%
\bibitem [{\citenamefont {Zhang}\ \emph {et~al.}(2014)\citenamefont {Zhang},
  \citenamefont {Zou}, \citenamefont {Jiang},\ and\ \citenamefont
  {Tang}}]{zhang2014strongly}%
  \BibitemOpen
  \bibfield  {author} {\bibinfo {author} {\bibfnamefont {X.}~\bibnamefont
  {Zhang}}, \bibinfo {author} {\bibfnamefont {C.-L.}\ \bibnamefont {Zou}},
  \bibinfo {author} {\bibfnamefont {L.}~\bibnamefont {Jiang}},\ and\ \bibinfo
  {author} {\bibfnamefont {H.~X.}\ \bibnamefont {Tang}},\ }\bibfield  {title}
  {\bibinfo {title} {\textit{Strongly coupled magnons and cavity microwave
  photons}},\ }\href {https://doi.org/10.1103/PhysRevLett.113.156401}
  {\bibfield  {journal} {\bibinfo  {journal} {Phys. Rev. Lett.}\ }\textbf
  {\bibinfo {volume} {113}},\ \bibinfo {pages} {156401} (\bibinfo {year}
  {2014})}\BibitemShut {NoStop}%
\bibitem [{\citenamefont {Tabuchi}\ \emph {et~al.}(2014)\citenamefont
  {Tabuchi}, \citenamefont {Ishino}, \citenamefont {Ishikawa}, \citenamefont
  {Yamazaki}, \citenamefont {Usami},\ and\ \citenamefont
  {Nakamura}}]{tabuchi2014hybridizing}%
  \BibitemOpen
  \bibfield  {author} {\bibinfo {author} {\bibfnamefont {Y.}~\bibnamefont
  {Tabuchi}}, \bibinfo {author} {\bibfnamefont {S.}~\bibnamefont {Ishino}},
  \bibinfo {author} {\bibfnamefont {T.}~\bibnamefont {Ishikawa}}, \bibinfo
  {author} {\bibfnamefont {R.}~\bibnamefont {Yamazaki}}, \bibinfo {author}
  {\bibfnamefont {K.}~\bibnamefont {Usami}},\ and\ \bibinfo {author}
  {\bibfnamefont {Y.}~\bibnamefont {Nakamura}},\ }\bibfield  {title} {\bibinfo
  {title} {\textit{Hybridizing ferromagnetic magnons and microwave photons in
  the quantum limit}},\ }\href {https://doi.org/10.1103/PhysRevLett.113.083603}
  {\bibfield  {journal} {\bibinfo  {journal} {Phys. Rev. Lett.}\ }\textbf
  {\bibinfo {volume} {113}},\ \bibinfo {pages} {083603} (\bibinfo {year}
  {2014})}\BibitemShut {NoStop}%
\bibitem [{\citenamefont {Goryachev}\ \emph {et~al.}(2014)\citenamefont
  {Goryachev}, \citenamefont {Farr}, \citenamefont {Creedon}, \citenamefont
  {Fan}, \citenamefont {Kostylev},\ and\ \citenamefont
  {Tobar}}]{goryachev2014high}%
  \BibitemOpen
  \bibfield  {author} {\bibinfo {author} {\bibfnamefont {M.}~\bibnamefont
  {Goryachev}}, \bibinfo {author} {\bibfnamefont {W.~G.}\ \bibnamefont {Farr}},
  \bibinfo {author} {\bibfnamefont {D.~L.}\ \bibnamefont {Creedon}}, \bibinfo
  {author} {\bibfnamefont {Y.}~\bibnamefont {Fan}}, \bibinfo {author}
  {\bibfnamefont {M.}~\bibnamefont {Kostylev}},\ and\ \bibinfo {author}
  {\bibfnamefont {M.~E.}\ \bibnamefont {Tobar}},\ }\bibfield  {title} {\bibinfo
  {title} {\textit{High-cooperativity cavity QED with magnons at microwave
  frequencies}},\ }\href {https://doi.org/10.1103/PhysRevApplied.2.054002}
  {\bibfield  {journal} {\bibinfo  {journal} {Phys. Rev. Appl.}\ }\textbf
  {\bibinfo {volume} {2}},\ \bibinfo {pages} {054002} (\bibinfo {year}
  {2014})}\BibitemShut {NoStop}%
\bibitem [{\citenamefont {Potts}\ and\ \citenamefont
  {Davis}(2020)}]{potts2020strong}%
  \BibitemOpen
  \bibfield  {author} {\bibinfo {author} {\bibfnamefont {C.~A.}\ \bibnamefont
  {Potts}}\ and\ \bibinfo {author} {\bibfnamefont {J.~P.}\ \bibnamefont
  {Davis}},\ }\bibfield  {title} {\bibinfo {title} {\textit{Strong
  magnon--photon coupling within a tunable cryogenic microwave cavity}},\
  }\href {https://doi.org/10.1063/5.0015660} {\bibfield  {journal} {\bibinfo
  {journal} {Appl. Phys. Lett.}\ }\textbf {\bibinfo {volume} {116}},\ \bibinfo
  {pages} {263503} (\bibinfo {year} {2020})}\BibitemShut {NoStop}%
\bibitem [{\citenamefont {Lachance-Quirion}\ \emph {et~al.}(2020)\citenamefont
  {Lachance-Quirion}, \citenamefont {Wolski}, \citenamefont {Tabuchi},
  \citenamefont {Kono}, \citenamefont {Usami},\ and\ \citenamefont
  {Nakamura}}]{lachance2020entanglement}%
  \BibitemOpen
  \bibfield  {author} {\bibinfo {author} {\bibfnamefont {D.}~\bibnamefont
  {Lachance-Quirion}}, \bibinfo {author} {\bibfnamefont {S.~P.}\ \bibnamefont
  {Wolski}}, \bibinfo {author} {\bibfnamefont {Y.}~\bibnamefont {Tabuchi}},
  \bibinfo {author} {\bibfnamefont {S.}~\bibnamefont {Kono}}, \bibinfo {author}
  {\bibfnamefont {K.}~\bibnamefont {Usami}},\ and\ \bibinfo {author}
  {\bibfnamefont {Y.}~\bibnamefont {Nakamura}},\ }\bibfield  {title} {\bibinfo
  {title} {\textit{Entanglement-based single-shot detection of a single magnon
  with a superconducting qubit}},\ }\href
  {https://doi.org/10.1126/science.aaz9236} {\bibfield  {journal} {\bibinfo
  {journal} {Science}\ }\textbf {\bibinfo {volume} {367}},\ \bibinfo {pages}
  {425} (\bibinfo {year} {2020})}\BibitemShut {NoStop}%
\bibitem [{\citenamefont {Lachance-Quirion}\ \emph {et~al.}(2017)\citenamefont
  {Lachance-Quirion}, \citenamefont {Tabuchi}, \citenamefont {Ishino},
  \citenamefont {Noguchi}, \citenamefont {Ishikawa}, \citenamefont {Yamazaki},\
  and\ \citenamefont {Nakamura}}]{lachance2017resolving}%
  \BibitemOpen
  \bibfield  {author} {\bibinfo {author} {\bibfnamefont {D.}~\bibnamefont
  {Lachance-Quirion}}, \bibinfo {author} {\bibfnamefont {Y.}~\bibnamefont
  {Tabuchi}}, \bibinfo {author} {\bibfnamefont {S.}~\bibnamefont {Ishino}},
  \bibinfo {author} {\bibfnamefont {A.}~\bibnamefont {Noguchi}}, \bibinfo
  {author} {\bibfnamefont {T.}~\bibnamefont {Ishikawa}}, \bibinfo {author}
  {\bibfnamefont {R.}~\bibnamefont {Yamazaki}},\ and\ \bibinfo {author}
  {\bibfnamefont {Y.}~\bibnamefont {Nakamura}},\ }\bibfield  {title} {\bibinfo
  {title} {\textit{Resolving quanta of collective spin excitations in a
  millimeter-sized ferromagnet}},\ }\href
  {https://doi.org/10.1126/sciadv.1603150} {\bibfield  {journal} {\bibinfo
  {journal} {Sci. Adv.}\ }\textbf {\bibinfo {volume} {3}},\ \bibinfo {pages}
  {e1603150} (\bibinfo {year} {2017})}\BibitemShut {NoStop}%
\bibitem [{\citenamefont {Tabuchi}\ \emph {et~al.}(2015)\citenamefont
  {Tabuchi}, \citenamefont {Ishino}, \citenamefont {Noguchi}, \citenamefont
  {Ishikawa}, \citenamefont {Yamazaki}, \citenamefont {Usami},\ and\
  \citenamefont {Nakamura}}]{tabuchi2015coherent}%
  \BibitemOpen
  \bibfield  {author} {\bibinfo {author} {\bibfnamefont {Y.}~\bibnamefont
  {Tabuchi}}, \bibinfo {author} {\bibfnamefont {S.}~\bibnamefont {Ishino}},
  \bibinfo {author} {\bibfnamefont {A.}~\bibnamefont {Noguchi}}, \bibinfo
  {author} {\bibfnamefont {T.}~\bibnamefont {Ishikawa}}, \bibinfo {author}
  {\bibfnamefont {R.}~\bibnamefont {Yamazaki}}, \bibinfo {author}
  {\bibfnamefont {K.}~\bibnamefont {Usami}},\ and\ \bibinfo {author}
  {\bibfnamefont {Y.}~\bibnamefont {Nakamura}},\ }\bibfield  {title} {\bibinfo
  {title} {\textit{Coherent coupling between a ferromagnetic magnon and a
  superconducting qubit}},\ }\href {https://doi.org/10.1126/science.aaa3693}
  {\bibfield  {journal} {\bibinfo  {journal} {Science}\ }\textbf {\bibinfo
  {volume} {349}},\ \bibinfo {pages} {405} (\bibinfo {year}
  {2015})}\BibitemShut {NoStop}%
\bibitem [{\citenamefont {Tabuchi}\ \emph {et~al.}(2016)\citenamefont
  {Tabuchi}, \citenamefont {Ishino}, \citenamefont {Noguchi}, \citenamefont
  {Ishikawa}, \citenamefont {Yamazaki}, \citenamefont {Usami},\ and\
  \citenamefont {Nakamura}}]{tabuchi2016quantum}%
  \BibitemOpen
  \bibfield  {author} {\bibinfo {author} {\bibfnamefont {Y.}~\bibnamefont
  {Tabuchi}}, \bibinfo {author} {\bibfnamefont {S.}~\bibnamefont {Ishino}},
  \bibinfo {author} {\bibfnamefont {A.}~\bibnamefont {Noguchi}}, \bibinfo
  {author} {\bibfnamefont {T.}~\bibnamefont {Ishikawa}}, \bibinfo {author}
  {\bibfnamefont {R.}~\bibnamefont {Yamazaki}}, \bibinfo {author}
  {\bibfnamefont {K.}~\bibnamefont {Usami}},\ and\ \bibinfo {author}
  {\bibfnamefont {Y.}~\bibnamefont {Nakamura}},\ }\bibfield  {title} {\bibinfo
  {title} {\textit{Quantum magnonics: The magnon meets the superconducting
  qubit}},\ }\href {https://doi.org/10.1016/j.crhy.2016.07.009} {\bibfield
  {journal} {\bibinfo  {journal} {C. R. Phys.}\ }\textbf {\bibinfo {volume}
  {17}},\ \bibinfo {pages} {729} (\bibinfo {year} {2016})}\BibitemShut
  {NoStop}%
\bibitem [{\citenamefont {Hisatomi}\ \emph {et~al.}(2016)\citenamefont
  {Hisatomi}, \citenamefont {Osada}, \citenamefont {Tabuchi}, \citenamefont
  {Ishikawa}, \citenamefont {Noguchi}, \citenamefont {Yamazaki}, \citenamefont
  {Usami},\ and\ \citenamefont {Nakamura}}]{hisatomi2016bidirectional}%
  \BibitemOpen
  \bibfield  {author} {\bibinfo {author} {\bibfnamefont {R.}~\bibnamefont
  {Hisatomi}}, \bibinfo {author} {\bibfnamefont {A.}~\bibnamefont {Osada}},
  \bibinfo {author} {\bibfnamefont {Y.}~\bibnamefont {Tabuchi}}, \bibinfo
  {author} {\bibfnamefont {T.}~\bibnamefont {Ishikawa}}, \bibinfo {author}
  {\bibfnamefont {A.}~\bibnamefont {Noguchi}}, \bibinfo {author} {\bibfnamefont
  {R.}~\bibnamefont {Yamazaki}}, \bibinfo {author} {\bibfnamefont
  {K.}~\bibnamefont {Usami}},\ and\ \bibinfo {author} {\bibfnamefont
  {Y.}~\bibnamefont {Nakamura}},\ }\bibfield  {title} {\bibinfo {title}
  {\textit{Bidirectional conversion between microwave and light via
  ferromagnetic magnons}},\ }\href {https://doi.org/10.1103/PhysRevB.93.174427}
  {\bibfield  {journal} {\bibinfo  {journal} {Phys. Rev. B}\ }\textbf {\bibinfo
  {volume} {93}},\ \bibinfo {pages} {174427} (\bibinfo {year}
  {2016})}\BibitemShut {NoStop}%
\bibitem [{\citenamefont {Zhu}\ \emph {et~al.}(2020)\citenamefont {Zhu},
  \citenamefont {Zhang}, \citenamefont {Han}, \citenamefont {Zou},
  \citenamefont {Zhong}, \citenamefont {Wang}, \citenamefont {Jiang},\ and\
  \citenamefont {Tang}}]{zhu2020waveguide}%
  \BibitemOpen
  \bibfield  {author} {\bibinfo {author} {\bibfnamefont {N.}~\bibnamefont
  {Zhu}}, \bibinfo {author} {\bibfnamefont {X.}~\bibnamefont {Zhang}}, \bibinfo
  {author} {\bibfnamefont {X.}~\bibnamefont {Han}}, \bibinfo {author}
  {\bibfnamefont {C.-L.}\ \bibnamefont {Zou}}, \bibinfo {author} {\bibfnamefont
  {C.}~\bibnamefont {Zhong}}, \bibinfo {author} {\bibfnamefont {C.-H.}\
  \bibnamefont {Wang}}, \bibinfo {author} {\bibfnamefont {L.}~\bibnamefont
  {Jiang}},\ and\ \bibinfo {author} {\bibfnamefont {H.~X.}\ \bibnamefont
  {Tang}},\ }\bibfield  {title} {\bibinfo {title} {\textit{Waveguide cavity
  optomagnonics for microwave-to-optics conversion}},\ }\href
  {https://doi.org/10.1364/OPTICA.397967} {\bibfield  {journal} {\bibinfo
  {journal} {Optica}\ }\textbf {\bibinfo {volume} {7}},\ \bibinfo {pages}
  {1291} (\bibinfo {year} {2020})}\BibitemShut {NoStop}%
\bibitem [{\citenamefont {Xu}\ \emph {et~al.}(2020)\citenamefont {Xu},
  \citenamefont {Zhong}, \citenamefont {Han}, \citenamefont {Jin},
  \citenamefont {Jiang},\ and\ \citenamefont {Zhang}}]{xu2020floquet}%
  \BibitemOpen
  \bibfield  {author} {\bibinfo {author} {\bibfnamefont {J.}~\bibnamefont
  {Xu}}, \bibinfo {author} {\bibfnamefont {C.}~\bibnamefont {Zhong}}, \bibinfo
  {author} {\bibfnamefont {X.}~\bibnamefont {Han}}, \bibinfo {author}
  {\bibfnamefont {D.}~\bibnamefont {Jin}}, \bibinfo {author} {\bibfnamefont
  {L.}~\bibnamefont {Jiang}},\ and\ \bibinfo {author} {\bibfnamefont
  {X.}~\bibnamefont {Zhang}},\ }\bibfield  {title} {\bibinfo {title}
  {\textit{Floquet Cavity Electromagnonics}},\ }\href
  {https://doi.org/10.1103/PhysRevLett.125.237201} {\bibfield  {journal}
  {\bibinfo  {journal} {Phys. Rev. Lett.}\ }\textbf {\bibinfo {volume} {125}},\
  \bibinfo {pages} {237201} (\bibinfo {year} {2020})}\BibitemShut {NoStop}%
\bibitem [{\citenamefont {Wang}\ \emph
  {et~al.}(2019{\natexlab{a}})\citenamefont {Wang}, \citenamefont {Rao},
  \citenamefont {Yang}, \citenamefont {Xu}, \citenamefont {Gui}, \citenamefont
  {Yao}, \citenamefont {You},\ and\ \citenamefont
  {Hu}}]{wang2019nonreciprocity}%
  \BibitemOpen
  \bibfield  {author} {\bibinfo {author} {\bibfnamefont {Y.-P.}\ \bibnamefont
  {Wang}}, \bibinfo {author} {\bibfnamefont {J.~W.}\ \bibnamefont {Rao}},
  \bibinfo {author} {\bibfnamefont {Y.}~\bibnamefont {Yang}}, \bibinfo {author}
  {\bibfnamefont {P.-C.}\ \bibnamefont {Xu}}, \bibinfo {author} {\bibfnamefont
  {Y.~S.}\ \bibnamefont {Gui}}, \bibinfo {author} {\bibfnamefont {B.~M.}\
  \bibnamefont {Yao}}, \bibinfo {author} {\bibfnamefont {J.~Q.}\ \bibnamefont
  {You}},\ and\ \bibinfo {author} {\bibfnamefont {C.-M.}\ \bibnamefont {Hu}},\
  }\bibfield  {title} {\bibinfo {title} {\textit{Nonreciprocity and
  unidirectional invisibility in cavity magnonics}},\ }\href
  {https://doi.org/10.1103/PhysRevLett.123.127202} {\bibfield  {journal}
  {\bibinfo  {journal} {Phys. Rev. Lett.}\ }\textbf {\bibinfo {volume} {123}},\
  \bibinfo {pages} {127202} (\bibinfo {year} {2019}{\natexlab{a}})}\BibitemShut
  {NoStop}%
\bibitem [{\citenamefont {Zhang}\ \emph {et~al.}(2016)\citenamefont {Zhang},
  \citenamefont {Zou}, \citenamefont {Jiang},\ and\ \citenamefont
  {Tang}}]{zhang2016cavity}%
  \BibitemOpen
  \bibfield  {author} {\bibinfo {author} {\bibfnamefont {X.}~\bibnamefont
  {Zhang}}, \bibinfo {author} {\bibfnamefont {C.-L.}\ \bibnamefont {Zou}},
  \bibinfo {author} {\bibfnamefont {L.}~\bibnamefont {Jiang}},\ and\ \bibinfo
  {author} {\bibfnamefont {H.~X.}\ \bibnamefont {Tang}},\ }\bibfield  {title}
  {\bibinfo {title} {\textit{Cavity magnomechanics}},\ }\href
  {https://doi.org/10.1126/sciadv.1501286} {\bibfield  {journal} {\bibinfo
  {journal} {Sci. Adv.}\ }\textbf {\bibinfo {volume} {2}},\ \bibinfo {pages}
  {e1501286} (\bibinfo {year} {2016})}\BibitemShut {NoStop}%
\bibitem [{\citenamefont {Cheng}\ \emph {et~al.}(2021)\citenamefont {Cheng},
  \citenamefont {Zhou}, \citenamefont {Peng}, \citenamefont {Kundu},
  \citenamefont {Li}, \citenamefont {Jin},\ and\ \citenamefont
  {Feng}}]{cheng2021tripartite}%
  \BibitemOpen
  \bibfield  {author} {\bibinfo {author} {\bibfnamefont {H.-J.}\ \bibnamefont
  {Cheng}}, \bibinfo {author} {\bibfnamefont {S.-J.}\ \bibnamefont {Zhou}},
  \bibinfo {author} {\bibfnamefont {J.-X.}\ \bibnamefont {Peng}}, \bibinfo
  {author} {\bibfnamefont {A.}~\bibnamefont {Kundu}}, \bibinfo {author}
  {\bibfnamefont {H.-X.}\ \bibnamefont {Li}}, \bibinfo {author} {\bibfnamefont
  {L.}~\bibnamefont {Jin}},\ and\ \bibinfo {author} {\bibfnamefont {X.-L.}\
  \bibnamefont {Feng}},\ }\bibfield  {title} {\bibinfo {title}
  {\textit{Tripartite entanglement in a Laguerre--Gaussian rotational-cavity
  system with an yttrium iron garnet sphere}},\ }\href
  {https://doi.org/10.1364/JOSAB.405097} {\bibfield  {journal} {\bibinfo
  {journal} {J. Opt. Soc. Am. B}\ }\textbf {\bibinfo {volume} {38}},\ \bibinfo
  {pages} {285} (\bibinfo {year} {2021})}\BibitemShut {NoStop}%
\bibitem [{\citenamefont {Nair}\ and\ \citenamefont
  {Agarwal}(2020)}]{nair2020deterministic}%
  \BibitemOpen
  \bibfield  {author} {\bibinfo {author} {\bibfnamefont {J.~M.}\ \bibnamefont
  {Nair}}\ and\ \bibinfo {author} {\bibfnamefont {G.}~\bibnamefont {Agarwal}},\
  }\bibfield  {title} {\bibinfo {title} {\textit{Deterministic quantum
  entanglement between macroscopic ferrite samples}},\ }\href
  {https://doi.org/10.1063/5.0015195} {\bibfield  {journal} {\bibinfo
  {journal} {Appl. Phys. Lett.}\ }\textbf {\bibinfo {volume} {117}},\ \bibinfo
  {pages} {084001} (\bibinfo {year} {2020})}\BibitemShut {NoStop}%
\bibitem [{\citenamefont {Li}\ and\ \citenamefont
  {Groeblacher}(2021)}]{li2021entangling}%
  \BibitemOpen
  \bibfield  {author} {\bibinfo {author} {\bibfnamefont {J.}~\bibnamefont
  {Li}}\ and\ \bibinfo {author} {\bibfnamefont {S.}~\bibnamefont
  {Groeblacher}},\ }\bibfield  {title} {\bibinfo {title} {\textit{Entangling
  the vibrational modes of two massive ferromagnetic spheres using cavity
  magnomechanics}},\ }\href {https://doi.org/10.1088/2058-9565/abd982}
  {\bibfield  {journal} {\bibinfo  {journal} {Quantum Sci. Tech.}\ }\textbf
  {\bibinfo {volume} {6}},\ \bibinfo {pages} {024005} (\bibinfo {year}
  {2021})}\BibitemShut {NoStop}%
\bibitem [{\citenamefont {Li}\ and\ \citenamefont
  {Zhu}(2019)}]{li2019entangling}%
  \BibitemOpen
  \bibfield  {author} {\bibinfo {author} {\bibfnamefont {J.}~\bibnamefont
  {Li}}\ and\ \bibinfo {author} {\bibfnamefont {S.-Y.}\ \bibnamefont {Zhu}},\
  }\bibfield  {title} {\bibinfo {title} {\textit{Entangling two magnon modes
  via magnetostrictive interaction}},\ }\href
  {https://doi.org/10.1088/1367-2630/ab3508} {\bibfield  {journal} {\bibinfo
  {journal} {New J. Phys.}\ }\textbf {\bibinfo {volume} {21}},\ \bibinfo
  {pages} {085001} (\bibinfo {year} {2019})}\BibitemShut {NoStop}%
\bibitem [{\citenamefont {Yang}\ \emph
  {et~al.}(2020{\natexlab{a}})\citenamefont {Yang}, \citenamefont {Liu},
  \citenamefont {Zhu}, \citenamefont {Liu},\ and\ \citenamefont
  {Yang}}]{yang2020nonreciprocal}%
  \BibitemOpen
  \bibfield  {author} {\bibinfo {author} {\bibfnamefont {Z.-B.}\ \bibnamefont
  {Yang}}, \bibinfo {author} {\bibfnamefont {J.-S.}\ \bibnamefont {Liu}},
  \bibinfo {author} {\bibfnamefont {A.-D.}\ \bibnamefont {Zhu}}, \bibinfo
  {author} {\bibfnamefont {H.-Y.}\ \bibnamefont {Liu}},\ and\ \bibinfo {author}
  {\bibfnamefont {R.-C.}\ \bibnamefont {Yang}},\ }\bibfield  {title} {\bibinfo
  {title} {\textit{Nonreciprocal Transmission and Nonreciprocal Entanglement in
  a Spinning Microwave Magnomechanical System}},\ }\href
  {https://doi.org/10.1002/andp.202000196} {\bibfield  {journal} {\bibinfo
  {journal} {Ann. Phys.}\ }\textbf {\bibinfo {volume} {532}},\ \bibinfo {pages}
  {2000196} (\bibinfo {year} {2020}{\natexlab{a}})}\BibitemShut {NoStop}%
\bibitem [{\citenamefont {Li}\ \emph {et~al.}(2018)\citenamefont {Li},
  \citenamefont {Zhu},\ and\ \citenamefont {Agarwal}}]{li2018magnon}%
  \BibitemOpen
  \bibfield  {author} {\bibinfo {author} {\bibfnamefont {J.}~\bibnamefont
  {Li}}, \bibinfo {author} {\bibfnamefont {S.-Y.}\ \bibnamefont {Zhu}},\ and\
  \bibinfo {author} {\bibfnamefont {G.~S.}\ \bibnamefont {Agarwal}},\
  }\bibfield  {title} {\bibinfo {title} {\textit{Magnon-photon-phonon
  entanglement in cavity magnomechanics}},\ }\href
  {https://doi.org/10.1103/PhysRevLett.121.203601} {\bibfield  {journal}
  {\bibinfo  {journal} {Phys. Rev. Lett.}\ }\textbf {\bibinfo {volume} {121}},\
  \bibinfo {pages} {203601} (\bibinfo {year} {2018})}\BibitemShut {NoStop}%
\bibitem [{\citenamefont {Li}\ \emph {et~al.}(2021)\citenamefont {Li},
  \citenamefont {Wang}, \citenamefont {You},\ and\ \citenamefont
  {Zhu}}]{li2021squeezing}%
  \BibitemOpen
  \bibfield  {author} {\bibinfo {author} {\bibfnamefont {J.}~\bibnamefont
  {Li}}, \bibinfo {author} {\bibfnamefont {Y.-P.}\ \bibnamefont {Wang}},
  \bibinfo {author} {\bibfnamefont {J.~Q.}\ \bibnamefont {You}},\ and\ \bibinfo
  {author} {\bibfnamefont {S.-Y.}\ \bibnamefont {Zhu}},\ }\bibfield  {title}
  {\bibinfo {title} {\textit{Squeezing Microwave Fields via Magnetostrictive
  Interaction}},\ }\href@noop {} {\bibfield  {journal} {\bibinfo  {journal}
  {arXiv:2101.02796}\ } (\bibinfo {year} {2021})}\BibitemShut {NoStop}%
\bibitem [{\citenamefont {Li}\ \emph {et~al.}(2019)\citenamefont {Li},
  \citenamefont {Zhu},\ and\ \citenamefont {Agarwal}}]{li2019squeezed}%
  \BibitemOpen
  \bibfield  {author} {\bibinfo {author} {\bibfnamefont {J.}~\bibnamefont
  {Li}}, \bibinfo {author} {\bibfnamefont {S.-Y.}\ \bibnamefont {Zhu}},\ and\
  \bibinfo {author} {\bibfnamefont {G.~S.}\ \bibnamefont {Agarwal}},\
  }\bibfield  {title} {\bibinfo {title} {\textit{Squeezed states of magnons and
  phonons in cavity magnomechanics}},\ }\href
  {https://doi.org/10.1103/PhysRevA.99.021801} {\bibfield  {journal} {\bibinfo
  {journal} {Phys. Rev. A}\ }\textbf {\bibinfo {volume} {99}},\ \bibinfo
  {pages} {021801(R)} (\bibinfo {year} {2019})}\BibitemShut {NoStop}%
\bibitem [{\citenamefont {Zhang}\ \emph {et~al.}(2021)\citenamefont {Zhang},
  \citenamefont {Wang}, \citenamefont {Bai}, \citenamefont {Wang},
  \citenamefont {Zhang},\ and\ \citenamefont {Wang}}]{zhang2021generation}%
  \BibitemOpen
  \bibfield  {author} {\bibinfo {author} {\bibfnamefont {W.}~\bibnamefont
  {Zhang}}, \bibinfo {author} {\bibfnamefont {D.-Y.}\ \bibnamefont {Wang}},
  \bibinfo {author} {\bibfnamefont {C.-H.}\ \bibnamefont {Bai}}, \bibinfo
  {author} {\bibfnamefont {T.}~\bibnamefont {Wang}}, \bibinfo {author}
  {\bibfnamefont {S.}~\bibnamefont {Zhang}},\ and\ \bibinfo {author}
  {\bibfnamefont {H.-F.}\ \bibnamefont {Wang}},\ }\bibfield  {title} {\bibinfo
  {title} {\textit{Generation and transfer of squeezed states in a cavity
  magnomechanical system by two-tone microwave fields}},\ }\href
  {https://doi.org/10.1364/OE.418531} {\bibfield  {journal} {\bibinfo
  {journal} {Opt. Express}\ }\textbf {\bibinfo {volume} {29}},\ \bibinfo
  {pages} {11773} (\bibinfo {year} {2021})}\BibitemShut {NoStop}%
\bibitem [{\citenamefont {Sarma}\ \emph {et~al.}(2021)\citenamefont {Sarma},
  \citenamefont {Busch},\ and\ \citenamefont {Twamley}}]{sarma2021cavity}%
  \BibitemOpen
  \bibfield  {author} {\bibinfo {author} {\bibfnamefont {B.}~\bibnamefont
  {Sarma}}, \bibinfo {author} {\bibfnamefont {T.}~\bibnamefont {Busch}},\ and\
  \bibinfo {author} {\bibfnamefont {J.}~\bibnamefont {Twamley}},\ }\bibfield
  {title} {\bibinfo {title} {\textit{Cavity magnomechanical storage and
  retrieval of quantum states}},\ }\href@noop {} {\bibfield  {journal}
  {\bibinfo  {journal} {New J. Phys.}\ } (\bibinfo {year} {2021})}\BibitemShut
  {NoStop}%
\bibitem [{\citenamefont {Li}\ \emph {et~al.}(2020{\natexlab{b}})\citenamefont
  {Li}, \citenamefont {Yang}, \citenamefont {Shui}, \citenamefont {Li},
  \citenamefont {Wang},\ and\ \citenamefont {Wu}}]{li2020phase}%
  \BibitemOpen
  \bibfield  {author} {\bibinfo {author} {\bibfnamefont {X.}~\bibnamefont
  {Li}}, \bibinfo {author} {\bibfnamefont {W.-X.}\ \bibnamefont {Yang}},
  \bibinfo {author} {\bibfnamefont {T.}~\bibnamefont {Shui}}, \bibinfo {author}
  {\bibfnamefont {L.}~\bibnamefont {Li}}, \bibinfo {author} {\bibfnamefont
  {X.}~\bibnamefont {Wang}},\ and\ \bibinfo {author} {\bibfnamefont
  {Z.}~\bibnamefont {Wu}},\ }\bibfield  {title} {\bibinfo {title}
  {\textit{Phase control of the transmission in cavity magnomechanical system
  with magnon driving}},\ }\href {https://doi.org/10.1063/5.0028395} {\bibfield
   {journal} {\bibinfo  {journal} {J. Appl. Phys.}\ }\textbf {\bibinfo {volume}
  {128}},\ \bibinfo {pages} {233101} (\bibinfo {year}
  {2020}{\natexlab{b}})}\BibitemShut {NoStop}%
\bibitem [{\citenamefont {Yang}\ \emph {et~al.}(2021)\citenamefont {Yang},
  \citenamefont {Liu}, \citenamefont {Zhu}, \citenamefont {Liu},\ and\
  \citenamefont {Yang}}]{yang2021nonreciprocal}%
  \BibitemOpen
  \bibfield  {author} {\bibinfo {author} {\bibfnamefont {Z.-B.}\ \bibnamefont
  {Yang}}, \bibinfo {author} {\bibfnamefont {J.-S.}\ \bibnamefont {Liu}},
  \bibinfo {author} {\bibfnamefont {A.-D.}\ \bibnamefont {Zhu}}, \bibinfo
  {author} {\bibfnamefont {H.-Y.}\ \bibnamefont {Liu}},\ and\ \bibinfo {author}
  {\bibfnamefont {R.-C.}\ \bibnamefont {Yang}},\ }\bibfield  {title} {\bibinfo
  {title} {\textit{Nonreciprocal Transmission and Entanglement in a
  cavity-magnomechanical system}},\ }\href@noop {} {\bibfield  {journal}
  {\bibinfo  {journal} {arXiv:2101.09931}\ } (\bibinfo {year}
  {2021})}\BibitemShut {NoStop}%
\bibitem [{\citenamefont {Zhao}\ \emph {et~al.}(2021)\citenamefont {Zhao},
  \citenamefont {Wu}, \citenamefont {Li}, \citenamefont {Liu}, \citenamefont
  {Nori}, \citenamefont {Liu},\ and\ \citenamefont {Du}}]{zhao2021phase}%
  \BibitemOpen
  \bibfield  {author} {\bibinfo {author} {\bibfnamefont {J.}~\bibnamefont
  {Zhao}}, \bibinfo {author} {\bibfnamefont {L.}~\bibnamefont {Wu}}, \bibinfo
  {author} {\bibfnamefont {T.}~\bibnamefont {Li}}, \bibinfo {author}
  {\bibfnamefont {Y.-x.}\ \bibnamefont {Liu}}, \bibinfo {author} {\bibfnamefont
  {F.}~\bibnamefont {Nori}}, \bibinfo {author} {\bibfnamefont {Y.}~\bibnamefont
  {Liu}},\ and\ \bibinfo {author} {\bibfnamefont {J.}~\bibnamefont {Du}},\
  }\bibfield  {title} {\bibinfo {title} {\textit{Phase-controlled pathway
  interferences and switchable fast-slow light in a cavity-magnon polariton
  system}},\ }\href {https://doi.org/10.1103/PhysRevApplied.15.024056}
  {\bibfield  {journal} {\bibinfo  {journal} {Phys. Rev. Appl.}\ }\textbf
  {\bibinfo {volume} {15}},\ \bibinfo {pages} {024056} (\bibinfo {year}
  {2021})}\BibitemShut {NoStop}%
\bibitem [{\citenamefont {Kong}\ \emph {et~al.}(2019)\citenamefont {Kong},
  \citenamefont {Wang}, \citenamefont {Liu}, \citenamefont {Xiong},\ and\
  \citenamefont {Wu}}]{kong2019magnetically}%
  \BibitemOpen
  \bibfield  {author} {\bibinfo {author} {\bibfnamefont {C.}~\bibnamefont
  {Kong}}, \bibinfo {author} {\bibfnamefont {B.}~\bibnamefont {Wang}}, \bibinfo
  {author} {\bibfnamefont {Z.-X.}\ \bibnamefont {Liu}}, \bibinfo {author}
  {\bibfnamefont {H.}~\bibnamefont {Xiong}},\ and\ \bibinfo {author}
  {\bibfnamefont {Y.}~\bibnamefont {Wu}},\ }\bibfield  {title} {\bibinfo
  {title} {\textit{Magnetically controllable slow light based on
  magnetostrictive forces}},\ }\href {https://doi.org/10.1364/OE.27.005544}
  {\bibfield  {journal} {\bibinfo  {journal} {Opt. Express}\ }\textbf {\bibinfo
  {volume} {27}},\ \bibinfo {pages} {5544} (\bibinfo {year}
  {2019})}\BibitemShut {NoStop}%
\bibitem [{\citenamefont {Potts}\ \emph {et~al.}(2020)\citenamefont {Potts},
  \citenamefont {Bittencourt}, \citenamefont {{Viola Kusminskiy}},\ and\
  \citenamefont {Davis}}]{potts2020magnon}%
  \BibitemOpen
  \bibfield  {author} {\bibinfo {author} {\bibfnamefont {C.~A.}\ \bibnamefont
  {Potts}}, \bibinfo {author} {\bibfnamefont {V.~A. S.~V.}\ \bibnamefont
  {Bittencourt}}, \bibinfo {author} {\bibfnamefont {S.}~\bibnamefont {{Viola
  Kusminskiy}}},\ and\ \bibinfo {author} {\bibfnamefont {J.~P.}\ \bibnamefont
  {Davis}},\ }\bibfield  {title} {\bibinfo {title} {\textit{Magnon-Phonon
  Quantum Correlation Thermometry}},\ }\href
  {https://doi.org/10.1103/PhysRevApplied.13.064001} {\bibfield  {journal}
  {\bibinfo  {journal} {Phys. Rev. Appl.}\ }\textbf {\bibinfo {volume} {13}},\
  \bibinfo {pages} {064001} (\bibinfo {year} {2020})}\BibitemShut {NoStop}%
\bibitem [{\citenamefont {Ding}\ \emph {et~al.}(2021)\citenamefont {Ding},
  \citenamefont {Xin}, \citenamefont {Qin},\ and\ \citenamefont
  {Li}}]{ding2021enhanced}%
  \BibitemOpen
  \bibfield  {author} {\bibinfo {author} {\bibfnamefont {M.-S.}\ \bibnamefont
  {Ding}}, \bibinfo {author} {\bibfnamefont {X.-X.}\ \bibnamefont {Xin}},
  \bibinfo {author} {\bibfnamefont {S.-Y.}\ \bibnamefont {Qin}},\ and\ \bibinfo
  {author} {\bibfnamefont {C.}~\bibnamefont {Li}},\ }\bibfield  {title}
  {\bibinfo {title} {\textit{Enhanced entanglement and steering in PT-symmetric
  cavity magnomechanics}},\ }\href
  {https://doi.org/10.1016/j.optcom.2021.126903} {\bibfield  {journal}
  {\bibinfo  {journal} {Opt. Commun.}\ ,\ \bibinfo {pages} {126903}} (\bibinfo
  {year} {2021})}\BibitemShut {NoStop}%
\bibitem [{\citenamefont {Wang}\ \emph {et~al.}(2020)\citenamefont {Wang},
  \citenamefont {Yang}, \citenamefont {Liu}, \citenamefont {Bai}, \citenamefont
  {Wang}, \citenamefont {Zhang},\ and\ \citenamefont {Wang}}]{wang2020magnon}%
  \BibitemOpen
  \bibfield  {author} {\bibinfo {author} {\bibfnamefont {L.}~\bibnamefont
  {Wang}}, \bibinfo {author} {\bibfnamefont {Z.-X.}\ \bibnamefont {Yang}},
  \bibinfo {author} {\bibfnamefont {Y.-M.}\ \bibnamefont {Liu}}, \bibinfo
  {author} {\bibfnamefont {C.-H.}\ \bibnamefont {Bai}}, \bibinfo {author}
  {\bibfnamefont {D.-Y.}\ \bibnamefont {Wang}}, \bibinfo {author}
  {\bibfnamefont {S.}~\bibnamefont {Zhang}},\ and\ \bibinfo {author}
  {\bibfnamefont {H.-F.}\ \bibnamefont {Wang}},\ }\bibfield  {title} {\bibinfo
  {title} {\textit{Magnon Blockade in a PT-Symmetric-Like Cavity
  Magnomechanical System}},\ }\href {https://doi.org/10.1002/andp.202000028}
  {\bibfield  {journal} {\bibinfo  {journal} {Ann. Phys.}\ }\textbf {\bibinfo
  {volume} {532}},\ \bibinfo {pages} {2000028} (\bibinfo {year}
  {2020})}\BibitemShut {NoStop}%
\bibitem [{\citenamefont {Yang}\ \emph
  {et~al.}(2020{\natexlab{b}})\citenamefont {Yang}, \citenamefont {Wang},
  \citenamefont {Liu}, \citenamefont {Wang}, \citenamefont {Bai}, \citenamefont
  {Zhang},\ and\ \citenamefont {Wang}}]{yang2020ground}%
  \BibitemOpen
  \bibfield  {author} {\bibinfo {author} {\bibfnamefont {Z.-X.}\ \bibnamefont
  {Yang}}, \bibinfo {author} {\bibfnamefont {L.}~\bibnamefont {Wang}}, \bibinfo
  {author} {\bibfnamefont {Y.-M.}\ \bibnamefont {Liu}}, \bibinfo {author}
  {\bibfnamefont {D.-Y.}\ \bibnamefont {Wang}}, \bibinfo {author}
  {\bibfnamefont {C.-H.}\ \bibnamefont {Bai}}, \bibinfo {author} {\bibfnamefont
  {S.}~\bibnamefont {Zhang}},\ and\ \bibinfo {author} {\bibfnamefont {H.-F.}\
  \bibnamefont {Wang}},\ }\bibfield  {title} {\bibinfo {title} {\textit{Ground
  state cooling of magnomechanical resonator in PT-symmetric cavity
  magnomechanical system at room temperature}},\ }\href
  {https://doi.org/doi.org/10.1007/s11467-020-0996-y} {\bibfield  {journal}
  {\bibinfo  {journal} {Front. Phys.}\ }\textbf {\bibinfo {volume} {15}},\
  \bibinfo {pages} {1} (\bibinfo {year} {2020}{\natexlab{b}})}\BibitemShut
  {NoStop}%
\bibitem [{\citenamefont {Wang}\ \emph
  {et~al.}(2019{\natexlab{b}})\citenamefont {Wang}, \citenamefont {Zhang},
  \citenamefont {Li}, \citenamefont {Wu},\ and\ \citenamefont
  {Sun}}]{wang2019magnon}%
  \BibitemOpen
  \bibfield  {author} {\bibinfo {author} {\bibfnamefont {M.}~\bibnamefont
  {Wang}}, \bibinfo {author} {\bibfnamefont {D.}~\bibnamefont {Zhang}},
  \bibinfo {author} {\bibfnamefont {X.-H.}\ \bibnamefont {Li}}, \bibinfo
  {author} {\bibfnamefont {Y.-Y.}\ \bibnamefont {Wu}},\ and\ \bibinfo {author}
  {\bibfnamefont {Z.-Y.}\ \bibnamefont {Sun}},\ }\bibfield  {title} {\bibinfo
  {title} {\textit{Magnon chaos in PT-symmetric cavity magnomechanics}},\
  }\href {https://doi.org/10.1109/JPHOT.2019.2911963} {\bibfield  {journal}
  {\bibinfo  {journal} {IEEE Photonics J.}\ }\textbf {\bibinfo {volume} {11}},\
  \bibinfo {pages} {1} (\bibinfo {year} {2019}{\natexlab{b}})}\BibitemShut
  {NoStop}%
\bibitem [{\citenamefont {Mahboob}\ \emph {et~al.}(2013)\citenamefont
  {Mahboob}, \citenamefont {Nishiguchi}, \citenamefont {Fujiwara},\ and\
  \citenamefont {Yamaguchi}}]{mahboob2013phonon}%
  \BibitemOpen
  \bibfield  {author} {\bibinfo {author} {\bibfnamefont {I.}~\bibnamefont
  {Mahboob}}, \bibinfo {author} {\bibfnamefont {K.}~\bibnamefont {Nishiguchi}},
  \bibinfo {author} {\bibfnamefont {A.}~\bibnamefont {Fujiwara}},\ and\
  \bibinfo {author} {\bibfnamefont {H.}~\bibnamefont {Yamaguchi}},\ }\bibfield
  {title} {\bibinfo {title} {\textit{Phonon lasing in an electromechanical
  resonator}},\ }\href {https://doi.org/10.1103/PhysRevLett.110.127202}
  {\bibfield  {journal} {\bibinfo  {journal} {Phys. Rev. Lett.}\ }\textbf
  {\bibinfo {volume} {110}},\ \bibinfo {pages} {127202} (\bibinfo {year}
  {2013})}\BibitemShut {NoStop}%
\bibitem [{\citenamefont {Ding}\ \emph {et~al.}(2019)\citenamefont {Ding},
  \citenamefont {Zheng},\ and\ \citenamefont {Li}}]{ding2019phonon}%
  \BibitemOpen
  \bibfield  {author} {\bibinfo {author} {\bibfnamefont {M.-S.}\ \bibnamefont
  {Ding}}, \bibinfo {author} {\bibfnamefont {L.}~\bibnamefont {Zheng}},\ and\
  \bibinfo {author} {\bibfnamefont {C.}~\bibnamefont {Li}},\ }\bibfield
  {title} {\bibinfo {title} {\textit{Phonon laser in a cavity magnomechanical
  system}},\ }\href {https://doi.org/10.1038/s41598-019-52050-7} {\bibfield
  {journal} {\bibinfo  {journal} {Sci. Rep.}\ }\textbf {\bibinfo {volume}
  {9}},\ \bibinfo {pages} {15723} (\bibinfo {year} {2019})}\BibitemShut
  {NoStop}%
\bibitem [{\citenamefont {Rocheleau}\ \emph {et~al.}(2010)\citenamefont
  {Rocheleau}, \citenamefont {Ndukum}, \citenamefont {Macklin}, \citenamefont
  {Hertzberg}, \citenamefont {Clerk},\ and\ \citenamefont
  {Schwab}}]{rocheleau2010preparation}%
  \BibitemOpen
  \bibfield  {author} {\bibinfo {author} {\bibfnamefont {T.}~\bibnamefont
  {Rocheleau}}, \bibinfo {author} {\bibfnamefont {T.}~\bibnamefont {Ndukum}},
  \bibinfo {author} {\bibfnamefont {C.}~\bibnamefont {Macklin}}, \bibinfo
  {author} {\bibfnamefont {J.~B.}\ \bibnamefont {Hertzberg}}, \bibinfo {author}
  {\bibfnamefont {A.~A.}\ \bibnamefont {Clerk}},\ and\ \bibinfo {author}
  {\bibfnamefont {K.}~\bibnamefont {Schwab}},\ }\bibfield  {title} {\bibinfo
  {title} {\textit{Preparation and detection of a mechanical resonator near the
  ground state of motion}},\ }\href {https://doi.org/10.1038/nature08681}
  {\bibfield  {journal} {\bibinfo  {journal} {Nature}\ }\textbf {\bibinfo
  {volume} {463}},\ \bibinfo {pages} {72} (\bibinfo {year} {2010})}\BibitemShut
  {NoStop}%
\bibitem [{\citenamefont {Romero-Isart}(2011)}]{romero2011quantum}%
  \BibitemOpen
  \bibfield  {author} {\bibinfo {author} {\bibfnamefont {O.}~\bibnamefont
  {Romero-Isart}},\ }\bibfield  {title} {\bibinfo {title} {\textit{Quantum
  superposition of massive objects and collapse models}},\ }\href
  {https://doi.org/10.1103/PhysRevA.84.052121} {\bibfield  {journal} {\bibinfo
  {journal} {Phys. Rev. A}\ }\textbf {\bibinfo {volume} {84}},\ \bibinfo
  {pages} {052121} (\bibinfo {year} {2011})}\BibitemShut {NoStop}%
\bibitem [{\citenamefont {Romero-Isart}\ \emph {et~al.}(2011)\citenamefont
  {Romero-Isart}, \citenamefont {Pflanzer}, \citenamefont {Blaser},
  \citenamefont {Kaltenbaek}, \citenamefont {Kiesel}, \citenamefont
  {Aspelmeyer},\ and\ \citenamefont {Cirac}}]{romero2011large}%
  \BibitemOpen
  \bibfield  {author} {\bibinfo {author} {\bibfnamefont {O.}~\bibnamefont
  {Romero-Isart}}, \bibinfo {author} {\bibfnamefont {A.~C.}\ \bibnamefont
  {Pflanzer}}, \bibinfo {author} {\bibfnamefont {F.}~\bibnamefont {Blaser}},
  \bibinfo {author} {\bibfnamefont {R.}~\bibnamefont {Kaltenbaek}}, \bibinfo
  {author} {\bibfnamefont {N.}~\bibnamefont {Kiesel}}, \bibinfo {author}
  {\bibfnamefont {M.}~\bibnamefont {Aspelmeyer}},\ and\ \bibinfo {author}
  {\bibfnamefont {J.~I.}\ \bibnamefont {Cirac}},\ }\bibfield  {title} {\bibinfo
  {title} {\textit{Large quantum superpositions and interference of massive
  nanometer-sized objects}},\ }\href
  {https://doi.org/10.1103/PhysRevLett.107.020405} {\bibfield  {journal}
  {\bibinfo  {journal} {Phys. Rev. Lett.}\ }\textbf {\bibinfo {volume} {107}},\
  \bibinfo {pages} {020405} (\bibinfo {year} {2011})}\BibitemShut {NoStop}%
\bibitem [{\citenamefont {Childress}\ \emph {et~al.}(2017)\citenamefont
  {Childress}, \citenamefont {Schmidt}, \citenamefont {Kashkanova},
  \citenamefont {Brown}, \citenamefont {Harris}, \citenamefont {Aiello},
  \citenamefont {Marquardt},\ and\ \citenamefont
  {Harris}}]{childress2017cavity}%
  \BibitemOpen
  \bibfield  {author} {\bibinfo {author} {\bibfnamefont {L.}~\bibnamefont
  {Childress}}, \bibinfo {author} {\bibfnamefont {M.~P.}\ \bibnamefont
  {Schmidt}}, \bibinfo {author} {\bibfnamefont {A.~D.}\ \bibnamefont
  {Kashkanova}}, \bibinfo {author} {\bibfnamefont {C.~D.}\ \bibnamefont
  {Brown}}, \bibinfo {author} {\bibfnamefont {G.~I.}\ \bibnamefont {Harris}},
  \bibinfo {author} {\bibfnamefont {A.}~\bibnamefont {Aiello}}, \bibinfo
  {author} {\bibfnamefont {F.}~\bibnamefont {Marquardt}},\ and\ \bibinfo
  {author} {\bibfnamefont {J.~G.~E.}\ \bibnamefont {Harris}},\ }\bibfield
  {title} {\bibinfo {title} {\textit{Cavity optomechanics in a levitated helium
  drop}},\ }\href {https://doi.org/10.1103/PhysRevA.96.063842} {\bibfield
  {journal} {\bibinfo  {journal} {Phys. Rev. A}\ }\textbf {\bibinfo {volume}
  {96}},\ \bibinfo {pages} {063842} (\bibinfo {year} {2017})}\BibitemShut
  {NoStop}%
\bibitem [{\citenamefont {Deli{\'c}}\ \emph {et~al.}(2020)\citenamefont
  {Deli{\'c}}, \citenamefont {Reisenbauer}, \citenamefont {Dare}, \citenamefont
  {Grass}, \citenamefont {Vuleti{\'c}}, \citenamefont {Kiesel},\ and\
  \citenamefont {Aspelmeyer}}]{delic2020cooling}%
  \BibitemOpen
  \bibfield  {author} {\bibinfo {author} {\bibfnamefont {U.}~\bibnamefont
  {Deli{\'c}}}, \bibinfo {author} {\bibfnamefont {M.}~\bibnamefont
  {Reisenbauer}}, \bibinfo {author} {\bibfnamefont {K.}~\bibnamefont {Dare}},
  \bibinfo {author} {\bibfnamefont {D.}~\bibnamefont {Grass}}, \bibinfo
  {author} {\bibfnamefont {V.}~\bibnamefont {Vuleti{\'c}}}, \bibinfo {author}
  {\bibfnamefont {N.}~\bibnamefont {Kiesel}},\ and\ \bibinfo {author}
  {\bibfnamefont {M.}~\bibnamefont {Aspelmeyer}},\ }\bibfield  {title}
  {\bibinfo {title} {\textit{Cooling of a levitated nanoparticle to the
  motional quantum ground state}},\ }\href
  {https://doi.org/10.1126/science.aba3993} {\bibfield  {journal} {\bibinfo
  {journal} {Science}\ }\textbf {\bibinfo {volume} {367}},\ \bibinfo {pages}
  {892} (\bibinfo {year} {2020})}\BibitemShut {NoStop}%
\bibitem [{\citenamefont {Fletcher}\ and\ \citenamefont
  {Bell}(1959)}]{fletcher1959ferrimagnetic}%
  \BibitemOpen
  \bibfield  {author} {\bibinfo {author} {\bibfnamefont {P.~C.}\ \bibnamefont
  {Fletcher}}\ and\ \bibinfo {author} {\bibfnamefont {R.~O.}\ \bibnamefont
  {Bell}},\ }\bibfield  {title} {\bibinfo {title} {\textit{Ferrimagnetic
  resonance modes in spheres}},\ }\href {https://doi.org/10.1063/1.1735216}
  {\bibfield  {journal} {\bibinfo  {journal} {J. Appl. Phys}\ }\textbf
  {\bibinfo {volume} {30}},\ \bibinfo {pages} {687} (\bibinfo {year}
  {1959})}\BibitemShut {NoStop}%
\bibitem [{\citenamefont {Gilbert}(2004)}]{gilbert2004phenomenological}%
  \BibitemOpen
  \bibfield  {author} {\bibinfo {author} {\bibfnamefont {T.~L.}\ \bibnamefont
  {Gilbert}},\ }\bibfield  {title} {\bibinfo {title} {\textit{A
  phenomenological theory of damping in ferromagnetic materials}},\ }\href
  {https://doi.org/10.1109/TMAG.2004.836740} {\bibfield  {journal} {\bibinfo
  {journal} {IEEE T. Magn.}\ }\textbf {\bibinfo {volume} {40}},\ \bibinfo
  {pages} {3443} (\bibinfo {year} {2004})}\BibitemShut {NoStop}%
\bibitem [{\citenamefont {Nemarich}(1964)}]{nemarich1964contribution}%
  \BibitemOpen
  \bibfield  {author} {\bibinfo {author} {\bibfnamefont {J.}~\bibnamefont
  {Nemarich}},\ }\bibfield  {title} {\bibinfo {title} {\textit{Contribution of
  the Two-Magnon Process to Magnetostatic-Mode Relaxation}},\ }\href
  {https://doi.org/10.1103/PhysRev.136.A1657} {\bibfield  {journal} {\bibinfo
  {journal} {Phys. Rev.}\ }\textbf {\bibinfo {volume} {136}},\ \bibinfo {pages}
  {A1657} (\bibinfo {year} {1964})}\BibitemShut {NoStop}%
\bibitem [{\citenamefont {Klingler}\ \emph {et~al.}(2017)\citenamefont
  {Klingler}, \citenamefont {Maier-Flaig}, \citenamefont {Dubs}, \citenamefont
  {Surzhenko}, \citenamefont {Gross}, \citenamefont {Huebl}, \citenamefont
  {Goennenwein},\ and\ \citenamefont {Weiler}}]{klinger2017gilbertdamping}%
  \BibitemOpen
  \bibfield  {author} {\bibinfo {author} {\bibfnamefont {S.}~\bibnamefont
  {Klingler}}, \bibinfo {author} {\bibfnamefont {H.}~\bibnamefont
  {Maier-Flaig}}, \bibinfo {author} {\bibfnamefont {C.}~\bibnamefont {Dubs}},
  \bibinfo {author} {\bibfnamefont {O.}~\bibnamefont {Surzhenko}}, \bibinfo
  {author} {\bibfnamefont {R.}~\bibnamefont {Gross}}, \bibinfo {author}
  {\bibfnamefont {H.}~\bibnamefont {Huebl}}, \bibinfo {author} {\bibfnamefont
  {S.~T.~B.}\ \bibnamefont {Goennenwein}},\ and\ \bibinfo {author}
  {\bibfnamefont {M.}~\bibnamefont {Weiler}},\ }\bibfield  {title} {\bibinfo
  {title} {\textit{Gilbert damping of magnetostatic modes in a yttrium iron
  garnet sphere}},\ }\href {https://doi.org/10.1063/1.4977423} {\bibfield
  {journal} {\bibinfo  {journal} {Appl. Phys. Lett.}\ }\textbf {\bibinfo
  {volume} {110}},\ \bibinfo {pages} {092409} (\bibinfo {year}
  {2017})}\BibitemShut {NoStop}%
\bibitem [{\citenamefont {Keshtgar}\ \emph {et~al.}(2014)\citenamefont
  {Keshtgar}, \citenamefont {Zareyan},\ and\ \citenamefont
  {Bauer}}]{keshtgar2014acoustic}%
  \BibitemOpen
  \bibfield  {author} {\bibinfo {author} {\bibfnamefont {H.}~\bibnamefont
  {Keshtgar}}, \bibinfo {author} {\bibfnamefont {M.}~\bibnamefont {Zareyan}},\
  and\ \bibinfo {author} {\bibfnamefont {G.~E.}\ \bibnamefont {Bauer}},\
  }\bibfield  {title} {\bibinfo {title} {\textit{Acoustic parametric pumping of
  spin waves}},\ }\href {https://doi.org/10.1016/j.ssc.2013.12.026} {\bibfield
  {journal} {\bibinfo  {journal} {Solid State Commu.}\ }\textbf {\bibinfo
  {volume} {198}},\ \bibinfo {pages} {30} (\bibinfo {year} {2014})}\BibitemShut
  {NoStop}%
\bibitem [{\citenamefont {Callen}(1968)}]{callen1968magnetostriction}%
  \BibitemOpen
  \bibfield  {author} {\bibinfo {author} {\bibfnamefont {E.}~\bibnamefont
  {Callen}},\ }\bibfield  {title} {\bibinfo {title}
  {\textit{Magnetostriction}},\ }\href {https://doi.org/10.1063/1.2163507}
  {\bibfield  {journal} {\bibinfo  {journal} {J. Appl. Phys.}\ }\textbf
  {\bibinfo {volume} {39}},\ \bibinfo {pages} {519} (\bibinfo {year}
  {1968})}\BibitemShut {NoStop}%
\bibitem [{\citenamefont {Gurevich}\ and\ \citenamefont
  {Melkov}(1996)}]{gurevich1996magnetization}%
  \BibitemOpen
  \bibfield  {author} {\bibinfo {author} {\bibfnamefont {A.~C.}\ \bibnamefont
  {Gurevich}}\ and\ \bibinfo {author} {\bibfnamefont {G.~A.}\ \bibnamefont
  {Melkov}},\ }\href@noop {} {\emph {\bibinfo {title} {\textit{Magnetization
  Oscillations and Waves}}}},\ Vol.~\bibinfo {volume} {1}\ (\bibinfo
  {publisher} {CRC Press},\ \bibinfo {year} {1996})\BibitemShut {NoStop}%
\bibitem [{\citenamefont {Spencer}\ and\ \citenamefont
  {LeCraw}(1958)}]{spencer1958magneto}%
  \BibitemOpen
  \bibfield  {author} {\bibinfo {author} {\bibfnamefont {E.~G.}\ \bibnamefont
  {Spencer}}\ and\ \bibinfo {author} {\bibfnamefont {R.~C.}\ \bibnamefont
  {LeCraw}},\ }\bibfield  {title} {\bibinfo {title} {\textit{Magnetoacoustic
  resonance in yttrium iron garnet}},\ }\href
  {https://doi.org/10.1103/PhysRevLett.1.241} {\bibfield  {journal} {\bibinfo
  {journal} {Phys. Rev. Lett.}\ }\textbf {\bibinfo {volume} {1}},\ \bibinfo
  {pages} {241} (\bibinfo {year} {1958})}\BibitemShut {NoStop}%
\bibitem [{\citenamefont {LeCraw}\ \emph {et~al.}(1961)\citenamefont {LeCraw},
  \citenamefont {Spencer},\ and\ \citenamefont {Gordon}}]{lecraw1961extremely}%
  \BibitemOpen
  \bibfield  {author} {\bibinfo {author} {\bibfnamefont {R.~C.}\ \bibnamefont
  {LeCraw}}, \bibinfo {author} {\bibfnamefont {E.~G.}\ \bibnamefont
  {Spencer}},\ and\ \bibinfo {author} {\bibfnamefont {E.~I.}\ \bibnamefont
  {Gordon}},\ }\bibfield  {title} {\bibinfo {title} {\textit{Extremely low loss
  acoustic resonance in single-crystal garnet spheres}},\ }\href
  {https://doi.org/10.1103/PhysRevLett.6.620} {\bibfield  {journal} {\bibinfo
  {journal} {Phys. Rev. Lett.}\ }\textbf {\bibinfo {volume} {6}},\ \bibinfo
  {pages} {620} (\bibinfo {year} {1961})}\BibitemShut {NoStop}%
\bibitem [{Fer(2020)}]{FerriSphere}%
  \BibitemOpen
  \href@noop {} {}\bibinfo {howpublished} {See
  \url{http://www.ferrisphere.com/} for information on purchasing YIG spheres.}
  (\bibinfo {year} {2020}),\ \bibinfo {note} {accessed: 2020-05-05}\BibitemShut
  {NoStop}%
\bibitem [{\citenamefont {Kepesidis}\ \emph {et~al.}(2013)\citenamefont
  {Kepesidis}, \citenamefont {Bennett}, \citenamefont {Portolan}, \citenamefont
  {Lukin},\ and\ \citenamefont {Rabl}}]{kepesidis2013phonon}%
  \BibitemOpen
  \bibfield  {author} {\bibinfo {author} {\bibfnamefont {K.~V.}\ \bibnamefont
  {Kepesidis}}, \bibinfo {author} {\bibfnamefont {S.~D.}\ \bibnamefont
  {Bennett}}, \bibinfo {author} {\bibfnamefont {S.}~\bibnamefont {Portolan}},
  \bibinfo {author} {\bibfnamefont {M.~D.}\ \bibnamefont {Lukin}},\ and\
  \bibinfo {author} {\bibfnamefont {P.}~\bibnamefont {Rabl}},\ }\bibfield
  {title} {\bibinfo {title} {\textit{Phonon cooling and lasing with
  nitrogen-vacancy centers in diamond}},\ }\href
  {https://doi.org/10.1103/PhysRevB.88.064105} {\bibfield  {journal} {\bibinfo
  {journal} {Phys. Rev. B}\ }\textbf {\bibinfo {volume} {88}},\ \bibinfo
  {pages} {064105} (\bibinfo {year} {2013})}\BibitemShut {NoStop}%
\bibitem [{\citenamefont {Kemiktarak}\ \emph {et~al.}(2014)\citenamefont
  {Kemiktarak}, \citenamefont {Durand}, \citenamefont {Metcalfe},\ and\
  \citenamefont {Lawall}}]{kemiktarak2014mode}%
  \BibitemOpen
  \bibfield  {author} {\bibinfo {author} {\bibfnamefont {U.}~\bibnamefont
  {Kemiktarak}}, \bibinfo {author} {\bibfnamefont {M.}~\bibnamefont {Durand}},
  \bibinfo {author} {\bibfnamefont {M.}~\bibnamefont {Metcalfe}},\ and\
  \bibinfo {author} {\bibfnamefont {J.}~\bibnamefont {Lawall}},\ }\bibfield
  {title} {\bibinfo {title} {\textit{Mode competition and anomalous cooling in
  a multimode phonon laser}},\ }\href
  {https://doi.org/10.1103/PhysRevLett.113.030802} {\bibfield  {journal}
  {\bibinfo  {journal} {Phys. Rev. Lett.}\ }\textbf {\bibinfo {volume} {113}},\
  \bibinfo {pages} {030802} (\bibinfo {year} {2014})}\BibitemShut {NoStop}%
\bibitem [{\citenamefont {Wang}\ and\ \citenamefont
  {Hsu}(1970)}]{wang1970spin}%
  \BibitemOpen
  \bibfield  {author} {\bibinfo {author} {\bibfnamefont {S.}~\bibnamefont
  {Wang}}\ and\ \bibinfo {author} {\bibfnamefont {T.-l.}\ \bibnamefont {Hsu}},\
  }\bibfield  {title} {\bibinfo {title} {\textit{Spin-wave experiments:
  parametric excitation of acoustic waves and mode-locking of spin waves}},\
  }\href {https://doi.org/10.1063/1.1653115} {\bibfield  {journal} {\bibinfo
  {journal} {Appl. Phys. Lett.}\ }\textbf {\bibinfo {volume} {16}},\ \bibinfo
  {pages} {111} (\bibinfo {year} {1970})}\BibitemShut {NoStop}%
\bibitem [{\citenamefont {Vahala}\ \emph {et~al.}(2009)\citenamefont {Vahala},
  \citenamefont {Herrmann}, \citenamefont {Kn{\"u}nz}, \citenamefont
  {Batteiger}, \citenamefont {Saathoff}, \citenamefont {H{\"a}nsch},\ and\
  \citenamefont {Udem}}]{vahala2009phonon}%
  \BibitemOpen
  \bibfield  {author} {\bibinfo {author} {\bibfnamefont {K.}~\bibnamefont
  {Vahala}}, \bibinfo {author} {\bibfnamefont {M.}~\bibnamefont {Herrmann}},
  \bibinfo {author} {\bibfnamefont {S.}~\bibnamefont {Kn{\"u}nz}}, \bibinfo
  {author} {\bibfnamefont {V.}~\bibnamefont {Batteiger}}, \bibinfo {author}
  {\bibfnamefont {G.}~\bibnamefont {Saathoff}}, \bibinfo {author}
  {\bibfnamefont {T.}~\bibnamefont {H{\"a}nsch}},\ and\ \bibinfo {author}
  {\bibfnamefont {T.}~\bibnamefont {Udem}},\ }\bibfield  {title} {\bibinfo
  {title} {\textit{A phonon laser}},\ }\href
  {https://doi.org/10.1038/nphys1367} {\bibfield  {journal} {\bibinfo
  {journal} {Nat. Phys.}\ }\textbf {\bibinfo {volume} {5}},\ \bibinfo {pages}
  {682} (\bibinfo {year} {2009})}\BibitemShut {NoStop}%
\bibitem [{\citenamefont {Braginsky}(1988)}]{braginskiui1988resolution}%
  \BibitemOpen
  \bibfield  {author} {\bibinfo {author} {\bibfnamefont {V.~B.}\ \bibnamefont
  {Braginsky}},\ }\bibfield  {title} {\bibinfo {title} {\textit{Resolution in
  macroscopic measurements: progress and prospects}},\ }\href
  {https://doi.org/10.1070/pu1988v031n09abeh005622} {\bibfield  {journal}
  {\bibinfo  {journal} {Sov. Phys. Usp.}\ }\textbf {\bibinfo {volume} {31}},\
  \bibinfo {pages} {836} (\bibinfo {year} {1988})}\BibitemShut {NoStop}%
\bibitem [{\citenamefont {Bushev}\ \emph {et~al.}(2006)\citenamefont {Bushev},
  \citenamefont {Rotter}, \citenamefont {Wilson}, \citenamefont {Dubin},
  \citenamefont {Becher}, \citenamefont {Eschner}, \citenamefont {Blatt},
  \citenamefont {Steixner}, \citenamefont {Rabl},\ and\ \citenamefont
  {Zoller}}]{bushev2006feedback}%
  \BibitemOpen
  \bibfield  {author} {\bibinfo {author} {\bibfnamefont {P.}~\bibnamefont
  {Bushev}}, \bibinfo {author} {\bibfnamefont {D.}~\bibnamefont {Rotter}},
  \bibinfo {author} {\bibfnamefont {A.}~\bibnamefont {Wilson}}, \bibinfo
  {author} {\bibfnamefont {F.}~\bibnamefont {Dubin}}, \bibinfo {author}
  {\bibfnamefont {C.}~\bibnamefont {Becher}}, \bibinfo {author} {\bibfnamefont
  {J.}~\bibnamefont {Eschner}}, \bibinfo {author} {\bibfnamefont
  {R.}~\bibnamefont {Blatt}}, \bibinfo {author} {\bibfnamefont
  {V.}~\bibnamefont {Steixner}}, \bibinfo {author} {\bibfnamefont
  {P.}~\bibnamefont {Rabl}},\ and\ \bibinfo {author} {\bibfnamefont
  {P.}~\bibnamefont {Zoller}},\ }\bibfield  {title} {\bibinfo {title}
  {\textit{Feedback cooling of a single trapped ion}},\ }\href
  {https://doi.org/10.1103/PhysRevLett.96.043003} {\bibfield  {journal}
  {\bibinfo  {journal} {Phys. Rev. Lett.}\ }\textbf {\bibinfo {volume} {96}},\
  \bibinfo {pages} {043003} (\bibinfo {year} {2006})}\BibitemShut {NoStop}%
\bibitem [{\citenamefont {Poggio}\ \emph {et~al.}(2007)\citenamefont {Poggio},
  \citenamefont {Degen}, \citenamefont {Mamin},\ and\ \citenamefont
  {Rugar}}]{poggio2007feedback}%
  \BibitemOpen
  \bibfield  {author} {\bibinfo {author} {\bibfnamefont {M.}~\bibnamefont
  {Poggio}}, \bibinfo {author} {\bibfnamefont {C.~L.}\ \bibnamefont {Degen}},
  \bibinfo {author} {\bibfnamefont {H.~J.}\ \bibnamefont {Mamin}},\ and\
  \bibinfo {author} {\bibfnamefont {D.}~\bibnamefont {Rugar}},\ }\bibfield
  {title} {\bibinfo {title} {\textit{Feedback cooling of a cantilever’s
  fundamental mode below 5 mK}},\ }\href
  {https://doi.org/10.1103/PhysRevLett.99.017201} {\bibfield  {journal}
  {\bibinfo  {journal} {Phys. Rev. Lett}\ }\textbf {\bibinfo {volume} {99}},\
  \bibinfo {pages} {017201} (\bibinfo {year} {2007})}\BibitemShut {NoStop}%
\bibitem [{\citenamefont {Clerk}\ \emph {et~al.}(2010)\citenamefont {Clerk},
  \citenamefont {Devoret}, \citenamefont {Girvin}, \citenamefont {Marquardt},\
  and\ \citenamefont {Schoelkopf}}]{clerk2010introductionto}%
  \BibitemOpen
  \bibfield  {author} {\bibinfo {author} {\bibfnamefont {A.~A.}\ \bibnamefont
  {Clerk}}, \bibinfo {author} {\bibfnamefont {M.~H.}\ \bibnamefont {Devoret}},
  \bibinfo {author} {\bibfnamefont {S.~M.}\ \bibnamefont {Girvin}}, \bibinfo
  {author} {\bibfnamefont {F.}~\bibnamefont {Marquardt}},\ and\ \bibinfo
  {author} {\bibfnamefont {R.~J.}\ \bibnamefont {Schoelkopf}},\ }\bibfield
  {title} {\bibinfo {title} {\textit{Introduction to quantum noise,
  measurement, and amplification}},\ }\href
  {https://doi.org/10.1103/RevModPhys.82.1155} {\bibfield  {journal} {\bibinfo
  {journal} {Rev. Mod. Phys.}\ }\textbf {\bibinfo {volume} {82}},\ \bibinfo
  {pages} {1155} (\bibinfo {year} {2010})}\BibitemShut {NoStop}%
\bibitem [{\citenamefont {Kotler}\ \emph {et~al.}(2021)\citenamefont {Kotler},
  \citenamefont {Peterson}, \citenamefont {Shojaee}, \citenamefont {Lecocq},
  \citenamefont {Cicak}, \citenamefont {Kwiatkowski}, \citenamefont {Geller},
  \citenamefont {Glancy}, \citenamefont {Knill}, \citenamefont {Simmonds} \emph
  {et~al.}}]{kotler2021direct}%
  \BibitemOpen
  \bibfield  {author} {\bibinfo {author} {\bibfnamefont {S.}~\bibnamefont
  {Kotler}}, \bibinfo {author} {\bibfnamefont {G.~A.}\ \bibnamefont
  {Peterson}}, \bibinfo {author} {\bibfnamefont {E.}~\bibnamefont {Shojaee}},
  \bibinfo {author} {\bibfnamefont {F.}~\bibnamefont {Lecocq}}, \bibinfo
  {author} {\bibfnamefont {K.}~\bibnamefont {Cicak}}, \bibinfo {author}
  {\bibfnamefont {A.}~\bibnamefont {Kwiatkowski}}, \bibinfo {author}
  {\bibfnamefont {S.}~\bibnamefont {Geller}}, \bibinfo {author} {\bibfnamefont
  {S.}~\bibnamefont {Glancy}}, \bibinfo {author} {\bibfnamefont
  {E.}~\bibnamefont {Knill}}, \bibinfo {author} {\bibfnamefont {R.~W.}\
  \bibnamefont {Simmonds}}, \emph {et~al.},\ }\bibfield  {title} {\bibinfo
  {title} {\textit{Direct observation of deterministic macroscopic
  entanglement}},\ }\href {https://doi.org/10.1126/science.abf2998} {\bibfield
  {journal} {\bibinfo  {journal} {Science}\ }\textbf {\bibinfo {volume}
  {372}},\ \bibinfo {pages} {622} (\bibinfo {year} {2021})}\BibitemShut
  {NoStop}%
\bibitem [{\citenamefont {Weis}\ \emph {et~al.}(2010)\citenamefont {Weis},
  \citenamefont {Rivi{\`e}re}, \citenamefont {Del{\'e}glise}, \citenamefont
  {Gavartin}, \citenamefont {Arcizet}, \citenamefont {Schliesser},\ and\
  \citenamefont {Kippenberg}}]{weis2010optomechanically}%
  \BibitemOpen
  \bibfield  {author} {\bibinfo {author} {\bibfnamefont {S.}~\bibnamefont
  {Weis}}, \bibinfo {author} {\bibfnamefont {R.}~\bibnamefont {Rivi{\`e}re}},
  \bibinfo {author} {\bibfnamefont {S.}~\bibnamefont {Del{\'e}glise}}, \bibinfo
  {author} {\bibfnamefont {E.}~\bibnamefont {Gavartin}}, \bibinfo {author}
  {\bibfnamefont {O.}~\bibnamefont {Arcizet}}, \bibinfo {author} {\bibfnamefont
  {A.}~\bibnamefont {Schliesser}},\ and\ \bibinfo {author} {\bibfnamefont
  {T.~J.}\ \bibnamefont {Kippenberg}},\ }\bibfield  {title} {\bibinfo {title}
  {\textit{Optomechanically induced transparency}},\ }\href
  {https://doi.org/10.1126/science.1195596} {\bibfield  {journal} {\bibinfo
  {journal} {Science}\ }\textbf {\bibinfo {volume} {330}},\ \bibinfo {pages}
  {1520} (\bibinfo {year} {2010})}\BibitemShut {NoStop}%
\bibitem [{\citenamefont {Safavi-Naeini}\ \emph {et~al.}(2011)\citenamefont
  {Safavi-Naeini}, \citenamefont {Alegre}, \citenamefont {Chan}, \citenamefont
  {Eichenfield}, \citenamefont {Winger}, \citenamefont {Lin}, \citenamefont
  {Hill}, \citenamefont {Chang},\ and\ \citenamefont
  {Painter}}]{safavi2011electromagnetically}%
  \BibitemOpen
  \bibfield  {author} {\bibinfo {author} {\bibfnamefont {A.~H.}\ \bibnamefont
  {Safavi-Naeini}}, \bibinfo {author} {\bibfnamefont {T.~M.}\ \bibnamefont
  {Alegre}}, \bibinfo {author} {\bibfnamefont {J.}~\bibnamefont {Chan}},
  \bibinfo {author} {\bibfnamefont {M.}~\bibnamefont {Eichenfield}}, \bibinfo
  {author} {\bibfnamefont {M.}~\bibnamefont {Winger}}, \bibinfo {author}
  {\bibfnamefont {Q.}~\bibnamefont {Lin}}, \bibinfo {author} {\bibfnamefont
  {J.~T.}\ \bibnamefont {Hill}}, \bibinfo {author} {\bibfnamefont {D.~E.}\
  \bibnamefont {Chang}},\ and\ \bibinfo {author} {\bibfnamefont
  {O.}~\bibnamefont {Painter}},\ }\bibfield  {title} {\bibinfo {title}
  {\textit{Electromagnetically induced transparency and slow light with
  optomechanics}},\ }\href {https://doi.org/10.1038/nature09933} {\bibfield
  {journal} {\bibinfo  {journal} {Nature}\ }\textbf {\bibinfo {volume} {472}},\
  \bibinfo {pages} {69} (\bibinfo {year} {2011})}\BibitemShut {NoStop}%
\bibitem [{\citenamefont {Wang}\ \emph {et~al.}(2018)\citenamefont {Wang},
  \citenamefont {Zhang}, \citenamefont {Zhang}, \citenamefont {Li},
  \citenamefont {Hu},\ and\ \citenamefont {You}}]{wang2018bistability}%
  \BibitemOpen
  \bibfield  {author} {\bibinfo {author} {\bibfnamefont {Y.-P.}\ \bibnamefont
  {Wang}}, \bibinfo {author} {\bibfnamefont {G.-Q.}\ \bibnamefont {Zhang}},
  \bibinfo {author} {\bibfnamefont {D.}~\bibnamefont {Zhang}}, \bibinfo
  {author} {\bibfnamefont {T.-F.}\ \bibnamefont {Li}}, \bibinfo {author}
  {\bibfnamefont {C.-M.}\ \bibnamefont {Hu}},\ and\ \bibinfo {author}
  {\bibfnamefont {J.~Q.}\ \bibnamefont {You}},\ }\bibfield  {title} {\bibinfo
  {title} {\textit{Bistability of cavity magnon polaritons}},\ }\href
  {https://doi.org/10.1103/PhysRevLett.120.057202} {\bibfield  {journal}
  {\bibinfo  {journal} {Phys. Rev. Lett.}\ }\textbf {\bibinfo {volume} {120}},\
  \bibinfo {pages} {057202} (\bibinfo {year} {2018})}\BibitemShut {NoStop}%
\bibitem [{\citenamefont {Gibbons}\ and\ \citenamefont
  {Chirba}(1958)}]{gibbons1958acoustical}%
  \BibitemOpen
  \bibfield  {author} {\bibinfo {author} {\bibfnamefont {D.~F.}\ \bibnamefont
  {Gibbons}}\ and\ \bibinfo {author} {\bibfnamefont {V.~G.}\ \bibnamefont
  {Chirba}},\ }\bibfield  {title} {\bibinfo {title} {\textit{Acoustical Loss
  and Young's Modulus of Yttrium Iron Garnet}},\ }\href
  {https://doi.org/10.1103/PhysRev.110.770} {\bibfield  {journal} {\bibinfo
  {journal} {Phys. Rev.}\ }\textbf {\bibinfo {volume} {110}},\ \bibinfo {pages}
  {770} (\bibinfo {year} {1958})}\BibitemShut {NoStop}%
\bibitem [{\citenamefont {Honda}\ and\ \citenamefont
  {Terada}(1907)}]{honda1905change}%
  \BibitemOpen
  \bibfield  {author} {\bibinfo {author} {\bibfnamefont {K.}~\bibnamefont
  {Honda}}\ and\ \bibinfo {author} {\bibfnamefont {T.}~\bibnamefont {Terada}},\
  }\bibfield  {title} {\bibinfo {title} {\textit{II. On the change of elastic
  constants of ferromagnetic substances by magnetization}},\ }\href
  {https://doi.org/10.1080/14786440709463584} {\bibfield  {journal} {\bibinfo
  {journal} {London Edinburgh Philos. Mag. J. Sci.}\ }\textbf {\bibinfo
  {volume} {13}},\ \bibinfo {pages} {36} (\bibinfo {year} {1907})}\BibitemShut
  {NoStop}%
\bibitem [{\citenamefont {Scheidler}\ \emph {et~al.}(2016)\citenamefont
  {Scheidler}, \citenamefont {Asnani},\ and\ \citenamefont
  {Dapino}}]{scheidler2016dynamically}%
  \BibitemOpen
  \bibfield  {author} {\bibinfo {author} {\bibfnamefont {J.~J.}\ \bibnamefont
  {Scheidler}}, \bibinfo {author} {\bibfnamefont {V.~M.}\ \bibnamefont
  {Asnani}},\ and\ \bibinfo {author} {\bibfnamefont {M.~J.}\ \bibnamefont
  {Dapino}},\ }\bibfield  {title} {\bibinfo {title} {\textit{Dynamically tuned
  magnetostrictive spring with electrically controlled stiffness}},\ }\href
  {https://doi.org/10.1088/0964-1726/25/3/035007} {\bibfield  {journal}
  {\bibinfo  {journal} {Smart Mater. Struct.}\ }\textbf {\bibinfo {volume}
  {25}},\ \bibinfo {pages} {035007} (\bibinfo {year} {2016})}\BibitemShut
  {NoStop}%
\bibitem [{\citenamefont {Hansen}\ \emph {et~al.}(1974)\citenamefont {Hansen},
  \citenamefont {R{\"o}schmann},\ and\ \citenamefont
  {Tolksdorf}}]{hansen1974saturation}%
  \BibitemOpen
  \bibfield  {author} {\bibinfo {author} {\bibfnamefont {P.}~\bibnamefont
  {Hansen}}, \bibinfo {author} {\bibfnamefont {P.}~\bibnamefont
  {R{\"o}schmann}},\ and\ \bibinfo {author} {\bibfnamefont {W.}~\bibnamefont
  {Tolksdorf}},\ }\bibfield  {title} {\bibinfo {title} {\textit{Saturation
  magnetization of gallium-substituted yttrium iron garnet}},\ }\href
  {https://doi.org/10.1063/1.1663657} {\bibfield  {journal} {\bibinfo
  {journal} {J. Appl. Phys.}\ }\textbf {\bibinfo {volume} {45}},\ \bibinfo
  {pages} {2728} (\bibinfo {year} {1974})}\BibitemShut {NoStop}%
\bibitem [{\citenamefont {Marquardt}\ \emph {et~al.}(2007)\citenamefont
  {Marquardt}, \citenamefont {Chen}, \citenamefont {Clerk},\ and\ \citenamefont
  {Girvin}}]{marquardt2007quantumtheory}%
  \BibitemOpen
  \bibfield  {author} {\bibinfo {author} {\bibfnamefont {F.}~\bibnamefont
  {Marquardt}}, \bibinfo {author} {\bibfnamefont {J.~P.}\ \bibnamefont {Chen}},
  \bibinfo {author} {\bibfnamefont {A.~A.}\ \bibnamefont {Clerk}},\ and\
  \bibinfo {author} {\bibfnamefont {S.~M.}\ \bibnamefont {Girvin}},\ }\bibfield
   {title} {\bibinfo {title} {\textit{Quantum Theory of Cavity-Assisted
  Sideband Cooling of Mechanical Motion}},\ }\href
  {https://doi.org/10.1103/PhysRevLett.99.093902} {\bibfield  {journal}
  {\bibinfo  {journal} {Phys. Rev. Lett.}\ }\textbf {\bibinfo {volume} {99}},\
  \bibinfo {pages} {093902} (\bibinfo {year} {2007})}\BibitemShut {NoStop}%
\bibitem [{\citenamefont {Safavi-Naeini}\ \emph {et~al.}(2013)\citenamefont
  {Safavi-Naeini}, \citenamefont {Chan}, \citenamefont {Hill}, \citenamefont
  {Gr\"{o}blacher}, \citenamefont {Miao}, \citenamefont {Chen}, \citenamefont
  {Aspelmeyer},\ and\ \citenamefont {Painter}}]{safavinaeini2013lasernoise}%
  \BibitemOpen
  \bibfield  {author} {\bibinfo {author} {\bibfnamefont {A.~H.}\ \bibnamefont
  {Safavi-Naeini}}, \bibinfo {author} {\bibfnamefont {J.}~\bibnamefont {Chan}},
  \bibinfo {author} {\bibfnamefont {J.~T.}\ \bibnamefont {Hill}}, \bibinfo
  {author} {\bibfnamefont {S.}~\bibnamefont {Gr\"{o}blacher}}, \bibinfo
  {author} {\bibfnamefont {H.}~\bibnamefont {Miao}}, \bibinfo {author}
  {\bibfnamefont {Y.}~\bibnamefont {Chen}}, \bibinfo {author} {\bibfnamefont
  {M.}~\bibnamefont {Aspelmeyer}},\ and\ \bibinfo {author} {\bibfnamefont
  {O.}~\bibnamefont {Painter}},\ }\bibfield  {title} {\bibinfo {title}
  {\textit{Laser noise in cavity-optomechanical cooling and thermometry}},\
  }\href {https://doi.org/10.1088/1367-2630/15/3/035007} {\bibfield  {journal}
  {\bibinfo  {journal} {New J. Phys.}\ }\textbf {\bibinfo {volume} {15}},\
  \bibinfo {pages} {035007} (\bibinfo {year} {2013})}\BibitemShut {NoStop}%
\end{thebibliography}%

\end{document}